\documentclass[tighten,twocolumn]{aastex63}

\usepackage{graphicx}
\usepackage{ulem}
\usepackage{amsmath}
\usepackage{wrapfig}
\usepackage[thinlines]{easytable}
\usepackage{tikz}
\usepackage{hyperref}
\usepackage{tabularx}
\usepackage{url}
\usepackage{lineno}

\graphicspath{{./}{figures/}}

\begin{document}


\title{The diverse outcomes of binary white dwarf mergers and connections to Galactic LISA sources}

\correspondingauthor{Kyle Kremer}
\email{kykremer@ucsd.edu}

\author[0000-0002-4086-3180]{Kyle Kremer}
\affiliation{University of California, San Diego, Department of Astronomy \& Astrophysics; La Jolla, CA 92093, USA}

\author[0000-0001-5228-6598]{Katelyn Breivik}
\affiliation{Carnegie Mellon University, McWilliams Center for Cosmology and Astrophysics, Department of Physics, Pittsburgh, PA 15213, USA}

\author[0000-0001-9582-881X]{Claire S. Ye}
\affil{Canadian Institute for Theoretical Astrophysics, University of Toronto, 60 St. George Street, Toronto, ON M5S 3H8, Canada}

\begin{abstract}
In the coming decade, the millihertz gravitational wave observatory LISA will provide the best constraints yet on the tens of thousands of close white dwarf binaries in the Milky Way, yielding unprecedented insights into the most abundant class of compact object binaries. Following inspiral via gravitational wave emission, interacting white dwarf binary pairs can lead to a multitude of outcomes, including AM Canum Venaticorum (AM CVn) binaries, R Coronae Borealis stars, young, rapidly-spinning single white dwarfs, (millisecond) magnetars, and a variety of explosive transients, most notably Type Ia supernovae. Current and future electromagnetic observations of these various outcomes coupled with the forthcoming flood of data from LISA place us on the precipice of a significant advance in our understanding of the long-term fate of white dwarf binaries. In this paper, we present a suite of mock catalogs of the Milky Way's white dwarf merger history, created using the population synthesis code \texttt{COSMIC} combined with a metallicity-dependent star formation history from FIRE-2 galaxy simulations. We summarize the various merger outcomes expected (based upon varying white dwarf masses and chemical compositions) and explore ways the rates of these outcomes may vary with model uncertainties pertaining to binary evolution. We publicly release these merger catalogs as a tool for facilitating connections between gravitational wave science and white dwarf binary astrophysics. 
\vspace{0.7cm}
\end{abstract}

\section{Introduction}
\label{sec:intro}

At present, the Milky Way is expected to host tens of millions of white dwarf binary pairs, by far the most abundant class of compact object binaries \citep[e.g.,][]{Nelemans2001a}. A subset of these systems are sufficiently compact (orbital separations of order $R_{\odot}$ or less) to enter Roche lobe contact on timescales of a Gyr or less via emission of gravitational waves \citep{Peters1964}. Since the 1980s, when \citet{IbenTutukov1984} and \citet{Webbink1984} first proposed massive white dwarf binary mergers as a mechanism for Type Ia supernovae (SN Ia), the broad astrophysical significance of these systems has been clear. Subsequent work has shown that the final stages of white dwarf binary evolution can lead to a diverse array of merger outcomes, determined by the combination of white dwarf masses, compositions, and merger dynamics. Shortly after the classic papers linking white dwarf binary mergers to SN Ia, \citet{SaioNomoto1985}
and \citet{NomotoIben1985} pointed out that in some cases off-center ignition of carbon triggered by accretion onto the more massive white dwarf may trigger conversion of the massive white dwarf to an oxygen-neon (ONe) composition, with subsequent electron captures ultimately leading to collapse into a neutron star \citep{Miyaji1980,Schwab2015}. More recent work studying the merger hydrodynamics \citep{Benz1990,Guillochon2010,Pakmor2010,Pakmor2012,Dan2011,Dan2014} and subsequent viscous \citep{Shen2012,Schwab2012} and thermal \citep{Schwab2016} evolution of white dwarf merger remnants have uncovered further complexity. These studies show that, beyond the canonical SN Ia and electron-capture collapse outcomes, off-center carbon ignition may lift the degeneracy of the accretor core in some cases, leading to a long-lived ($>10^3$yr) quasi-giant phase that may ultimately collapse into a massive white dwarf or lead to an iron-core collapse supernova and neutron star formation \citep[for an overview, see][]{Schwab2021}.

\begin{figure*}
    \centering
    \includegraphics[width=0.9\linewidth]{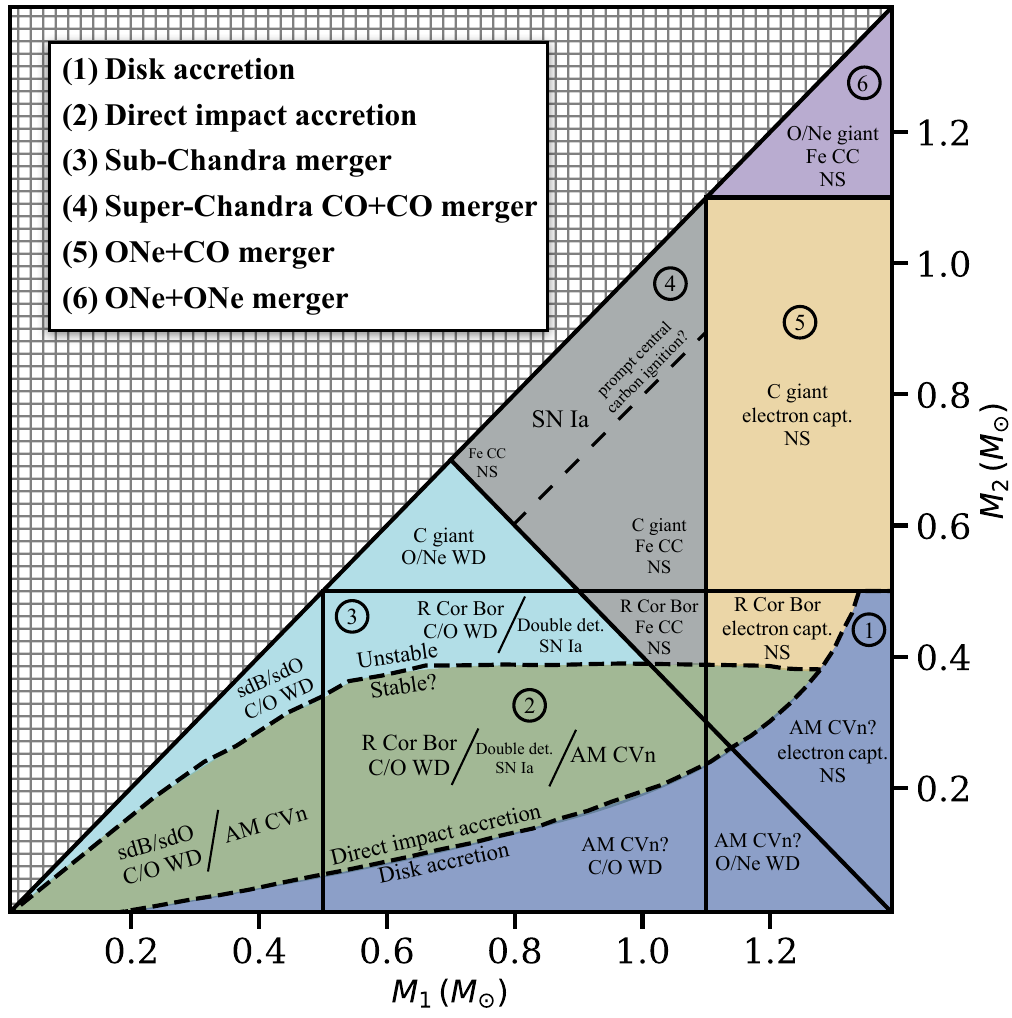}
    \caption{Summary of outcomes for interacting double white dwarf binaries for various component masses, $M_1$ and $M_2$. The solid black lines denote the approximate boundaries between different white dwarf compositions: He ($<0.5\,M_{\odot}$), CO ($0.5-1.1\,M_{\odot}$), and O/Ne ($>1.1\,M_{\odot}$). The diagonal solid line denotes the Chandrasekhar limit : $M_1+M_2=1.4\,M_{\odot}$. The two dashed curves mark the boundaries between disk and direct impact accretion (bottom curve) and dynamically stable versus unstable accretion \citep[upper curve, adopting the direct impact model of][]{Kremer2015}. The six colored regions denote boundaries of distinct interaction outcomes. \textbf{Region 1 (blue):} Disk accretion. May lead to long-lived AM CVn, although see \citet{Shen2015,Brown2016} for alternative. \textbf{Region 2 (green):} Direct impact accretion. If accretion is stable \citep[see][]{Marsh2004,Kremer2015} long-lived AM CVn is expected, if unstable, a merger occurs leading to an R~Cor~Bor star (or, in the case of two He white dwarfs, a core He-burning sdB or sdO star) and ultimately a CO white dwarf remnant. For CO accretor, a double detonation \citep[e.g.,][]{Shen2018} may also lead to instability, merger, and potentially a sub-Chandrasekhar SN Ia. \textbf{Region 3 (cyan):} Unstable mass transfer leads to formation of R~Cor~Bor star (C giant if both white dwarfs are CO composition) and ultimately a CO (ONe in latter case) white dwarf. \textbf{Region 4 (gray):} Merger of a super-Chandrasekhar CO white dwarf pair. If central ignition of carbon occurs \citep[likely for ``violent mergers" of near equal mass CO white dwarfs;][]{Pakmor2012,Dan2014} a SN Ia is expected. For lower mass mergers or highly asymmetric mergers, off-center carbon ignition lifts degeneracy of core, leading to formation of a C giant and ultimately collapse to a neutron star via iron core collapse \citep[Fe CC;][]{Schwab2021}. \textbf{Region 5 (gold):} Merger of CO and ONe pair leads to runaway electron captures in ONe core resulting in collapse to a neutron star \citep{Schwab2015}. \textbf{Region 6 (lavender):} Merger of ONe+ONe pair. Off-center ignition of oxygen leads to iron core collapse and formation of neutron star \citep{Shen2015}.}
    \label{fig:summary}
\end{figure*}

Meanwhile, less massive double white dwarf systems (total mass below the Chandrasekhar limit) have been invoked to explain objects like the R Coronae Borealis (R~Cor~Bor) stars and extreme helium stars \citep{Webbink1984,SaioJeffery2000,Clayton2007,Schwab2019}.\footnote{Giant-like objects similar to R~Cor~Bor stars could in principle also be formed by mergers of white dwarfs and non-degenerate stars, for example as the outcome of some cataclysmic variables \citep[e.g.,][]{Metzger2021}. We focus here exclusively on mergers involving pairs of white dwarfs.} Ultimately the cores of such R~Cor~Bor stars will ``burn out'' leaving behind a sub-Chandrasekhar carbon-oxygen (CO) or ONe white dwarf. Recent surveys such as the Sloan Digital Sky Survey (SDSS) and the Zwicky Transient Survey (ZTF) have identified samples of highly-magnetized ($B >10^6\,$G) and fast-rotating ($P_{\rm spin} \lesssim\,$hr) hot and young white dwarfs, suspected to have formed via recent sub-Chandrasekhar white dwarf mergers \citep{Ferrario2015,Caiazzo2021,Caiazzo2023}.

For binary mass ratio, $q$, near unity, unstable mass transfer and merger are all but inevitable upon Roche lobe contact \citep[e.g.,][]{Marsh2004}. However, for binaries with more asymmetric mass ratios, stable mass transfer may be possible, potentially connecting to the observed population of ultracompact AM Canum Venaticorum (AM CVn) systems \citep{Nather1981,Tutukov1996,Nelemans2001b,Kilic2014,Ramsay2018,vanRoestel2022}. For $q \lesssim 0.1$, an accretion disk forms enabling efficient redistribution of angular momentum back into the orbit, stabilizing the binary \citep[e.g.,][]{VerbuntRappaport1988}. For intermediate mass ratios ($q\sim 0.3$), the accretion flow can impact the surface of the accretor directly \citep{Webbink1984,Nelemans2001b}. In the absence of an extended disk, it may be more difficult for tides to stabilize the orbit, however some studies \citep{Marsh2004,Gokhale2007,Kremer2015} demonstrate that stable mass transfer may still be possible for some systems in this direct-impact regime.

\begin{deluxetable*}{l | p{8cm} | p{7.5cm}}
\tablecaption{Model variations\label{table:models}}
\tablehead{
\multicolumn{1}{l}{\textbf{Model}} &
\multicolumn{1}{l}{\textbf{\texttt{COSMIC} parameter change}} &
\multicolumn{1}{l}{\textbf{Binary evolution change}}
}
\startdata
Fiducial & None; $\alpha=1$, $q_c$ computed from \citet{Hurley2002,Claeys2014} \texttt{qcflag=2} in \texttt{inifile} & None \\
\hline
$\alpha0.25$ & $\alpha=0.25$ & Reduced CE efficiency; closer post-CE separations or failed CE (stellar merger) \\
\hline
$\alpha5$ & $\alpha=5$ & Increased CE efficiency; wider post-CE separations \\
\hline
$q3$ & $q_c=3$ for H-rich donors, following mass transfer prescriptions of \citet{Belczynski2008} \texttt{qcflag=4} in \texttt{inifile} & Increased critical mass ratio; allows stable mass transfer for more massive RLO donors \\
\enddata
\end{deluxetable*}

Alternatively, \citet{Shen2015} argues that even in cases where mass transfer is initially dynamically stable, nova-like outbursts triggered by helium ignition on the surface of the accretor lead to further inspiral and a dramatic increase in the mass transfer rate. In this case, \textit{all} interacting white dwarf binaries may merge, independent of their initial mass ratio. Under this assumption, AM CVn systems must form via channels alternative to the double degenerate scenario \citep[e.g.,][]{Savonije1986,IbenTutukov1987,Podsiadlowski2003}. \citet{Shen2015} notes that this may help resolve apparent discrepancies between the predicted and observed AM CVn space densities \citep{Nelemans2001b,Carter2013}. Observationally, this is further supported by the Extremely Low Mass (ELM) survey of white dwarfs \citep{Brown2016,Brown2020} which disfavors the stable mass transfer outcome for He+CO white dwarf mergers on the basis of observed AM CVn formation rates, but shows these mergers are consistent with observed R~Cor~Bor rates.

In some cases, detonation of helium on the surface of the accretor may trigger a second detonation within the accretor's carbon core \citep{Taam1980,Livne1990,ShenBildsten2009}. This double-detonation scenario may provide a mechanism for exploding sub-Chandrasekhar CO white dwarfs, enabling a pathway for sub-Chandrasekhar SN Ia \citep[e.g.,][]{ShenBildsten2014}. Recent observations of hypervelocity white dwarfs in \textit{Gaia} DR3 (presumably the surviving companions of exploded white dwarfs previously in sub-hour Roche-lobe filling orbits) have been linked to this double-degenerate detonation scenario \citep{Shen2018}. The detection of calcium signatures within nearby SN remnants may further point to a double-detonation scenario for sub-Chandrasekhar explosions \citep{Das2025}.

In Figure~\ref{fig:summary} we summarize the possible outcomes for interacting white dwarf binaries of varying masses and compositions. This figure is intended to be qualitative; the precise boundaries between the various outcomes (and in many cases the precise outcomes themselves) are uncertain. See the caption for further discussion and see \citet{Marsh2004,Dan2011,Dan2014,Shen2015,YungelsonKuranov2017,Shen2025} for similar schematics and discussion.

In the coming decade, the Laser Interferometer Space Antenna (LISA) will provide the most complete census yet of the close white dwarf binary population in the Milky Way via detection of gravitational waves emitted in the frequency range $10^{-5}-0.1\,$Hz \citep{LISA2023}. LISA is expected to observe tens of thousands of white dwarf binaries across the complete $M_1-M_2$ parameter space shown in Figure~\ref{fig:summary} \citep[e.g.,][]{Nelemans2001a,Breivik2020_WD,Thiele2023}. LISA will primarily observe inspiraling (pre-merger) white dwarf pairs. However, if a subset of low mass ratio white dwarf binaries undergo stable mass transfer, LISA may also observe a population of interacting white dwarf binaries. A subset of these may be observed as a unique class of ``reverse chirping'' gravitational wave sources evolving from higher to lower orbital frequencies \citep{Kremer2017}, and may also be multi-messenger sources \citep{Breivik2018}.

Binary population synthesis methods provide a valuable tool for understanding the formation and long-term outcomes of close white dwarf binaries in the Milky Way, including those present today and those that have merged throughout the Galaxy's history \citep{LipunovPostnov1988,TutukovYungleson1994,Yungleson1994,Han1995,Iben1997,Nelemans2001a,Nelemans2001b,Belczynski2005,Ruiter2010,Korol2017,YungelsonKuranov2017,Breivik2020_COSMIC,Thiele2023}. In this paper, we use the population synthesis code \texttt{COSMIC} to present a new suite of models of the Milky Way's white dwarf merger history to facilitate connections between white dwarf merger outcomes and the forthcoming LISA source catalog. Following \citet{Thiele2023}, these models adopt an empirically-derived metallicity-dependent binary fraction \citep[based on][]{Moe2019}, as well as a metallicity-dependent star formation history from the \texttt{m12i} galaxy of the Latte Suite of FIRE-2 simulations \citep{Wetzel2016,Hopkins2018} and stellar position assignments using the Ananke Framework \citep{Sanderson2020}.

This paper is organized as follows: In Section~\ref{sec:cosmic}, we summarize the methods implemented in our \texttt{COSMIC} models. In Section~\ref{sec:results}, we describe our predictions for white dwarf mergers across the Galactic history and the dependence of merger rates upon our binary evolution assumptions. In Section~\ref{sec:outcomes}, we discuss detection prospects of the various white dwarf merger outcomes. We summarize and conclude in Section~\ref{sec:summary}.

\section{Constructing Galactic White Dwarf Binary Populations}
\label{sec:cosmic}

\subsection{\texttt{COSMIC} population synthesis}
We use \texttt{COSMIC} \citep{Breivik2020_COSMIC} to simulate the Galactic population of white dwarf binaries that forms and evolves from the birth of the Galaxy up to the present day. Our simulation procedure closely follows the process described in \citet{Thiele2023}, but we include a brief description below. For a grid of fifteen metallicities ranging uniformly from $-2.3 < \log_{10}(Z/Z_{\odot}) < 0.18$, we initialize Zero Age Main Sequence (ZAMS) stellar populations with the same age and evolve them for $13.7\,\rm{Gyr}$ to capture all possible evolutionary outcomes within a Hubble Time. To create each population, we first sample stellar masses from a \citet{Kroupa2001} initial mass function. We then use the metallicity-dependent close binary fraction described in \citet{Moe2019} to probabilistically determine which stars have a companion and assign their orbital period. We sample secondary masses for binary systems from a uniform mass distribution with a minimum mass ratio of $q_{\rm{min}}=0.01$ and sample eccentricities from a uniform distribution between $0$ and $1$ \citep{Geller2019}. To apply proper statistical weights when generating our final astrophysical populations, we keep track of the total mass (including single and binary stars) that is initialized at ZAMS and the total number of double white dwarf binaries that form. 

Each initial population is evolved according to the public \texttt{COSMIC} models run as part of \citet{Thiele2023}. We summarize here the key features and parameters that are distinct from current \texttt{COSMIC} defaults. We assume that wind mass loss velocities follow the mass-dependent prescription defined in \citet{Belczynski2008} and vary the mass transfer stability and common envelope ejection efficiency assumptions in the variations described in Table~\ref{table:models}. We consider two mass transfer stability models. In our fiducial model, we follow \citet{Claeys2014} which applies the mass transfer assumptions from \citet{Hurley2002} for all stars except main sequence stars which have a reduced critical mass ratio of $M_{\rm{don}}/M_{\rm{acc}}=1.6$ (achieved by setting the \texttt{qcflag}=2 in the \texttt{COSMIC} \texttt{inifile}) and a common envelope ejection efficiency of $\alpha=1$. We further consider a mass transfer stability variation following \citet{Belczynski2008} which increases the critical mass ratio to $q_c=3$ for all hydrogen-rich donor stars, ultimately increasing the fraction of double white dwarf progenitors that undergo stable mass transfer (setting the \texttt{qcflag}=4 in the \texttt{COSMIC} \texttt{inifile}). We consider two variations in the ejection efficiency of the common envelope, $\alpha=[0.25, 5]$, which capture the wide range of uncertainty that applies to white dwarfs of different masses \citep[e.g.][]{demarco2009, zorotovic2010, ToonenNelemans2013,Camacho2014, scherbak2023, yamaguchi2024}.

\begin{figure*}
    \centering
    \includegraphics[width=\linewidth]{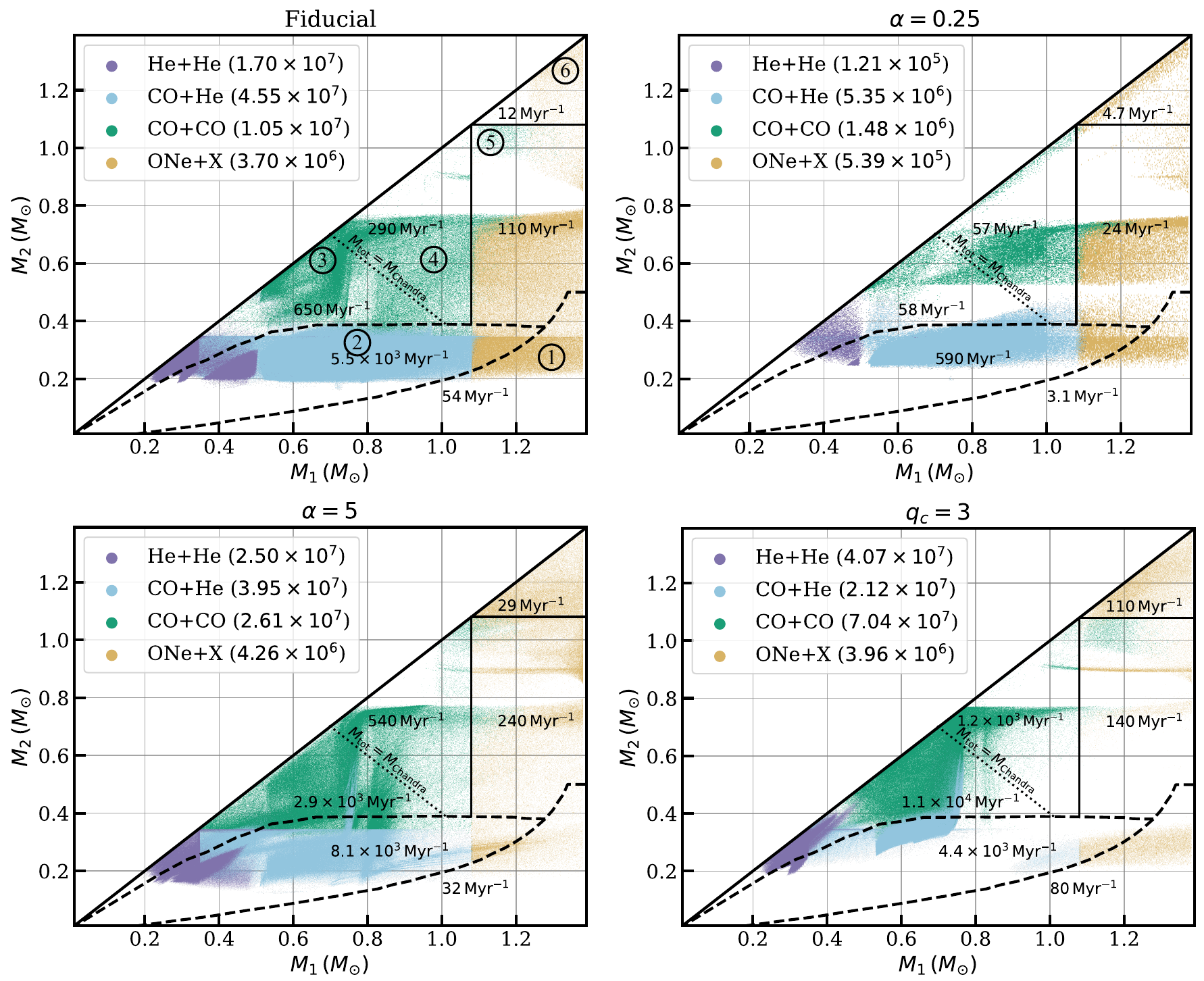}
    \caption{Secondary ($M_2$) versus primary ($M_1$) masses for all white dwarf mergers occurring throughout the full history of each of our Galactic population models. We mark the same six regions indicated in Figure~\ref{fig:summary} for distinct merger outcomes, and also indicate the Galactic merger rate within each region for events occurring within the past $100\,$Myr. Colors correspond to the four main white dwarf merger combinations, which are also roughly delineated by the black boundaries. The numbers in parentheses in each legend denote the total number of mergers of each white dwarf merger combination over the full Galactic history.
    }
    \label{fig:M1M2_all}
\end{figure*}

\begin{figure*}
    \centering
    \includegraphics[width=\linewidth]{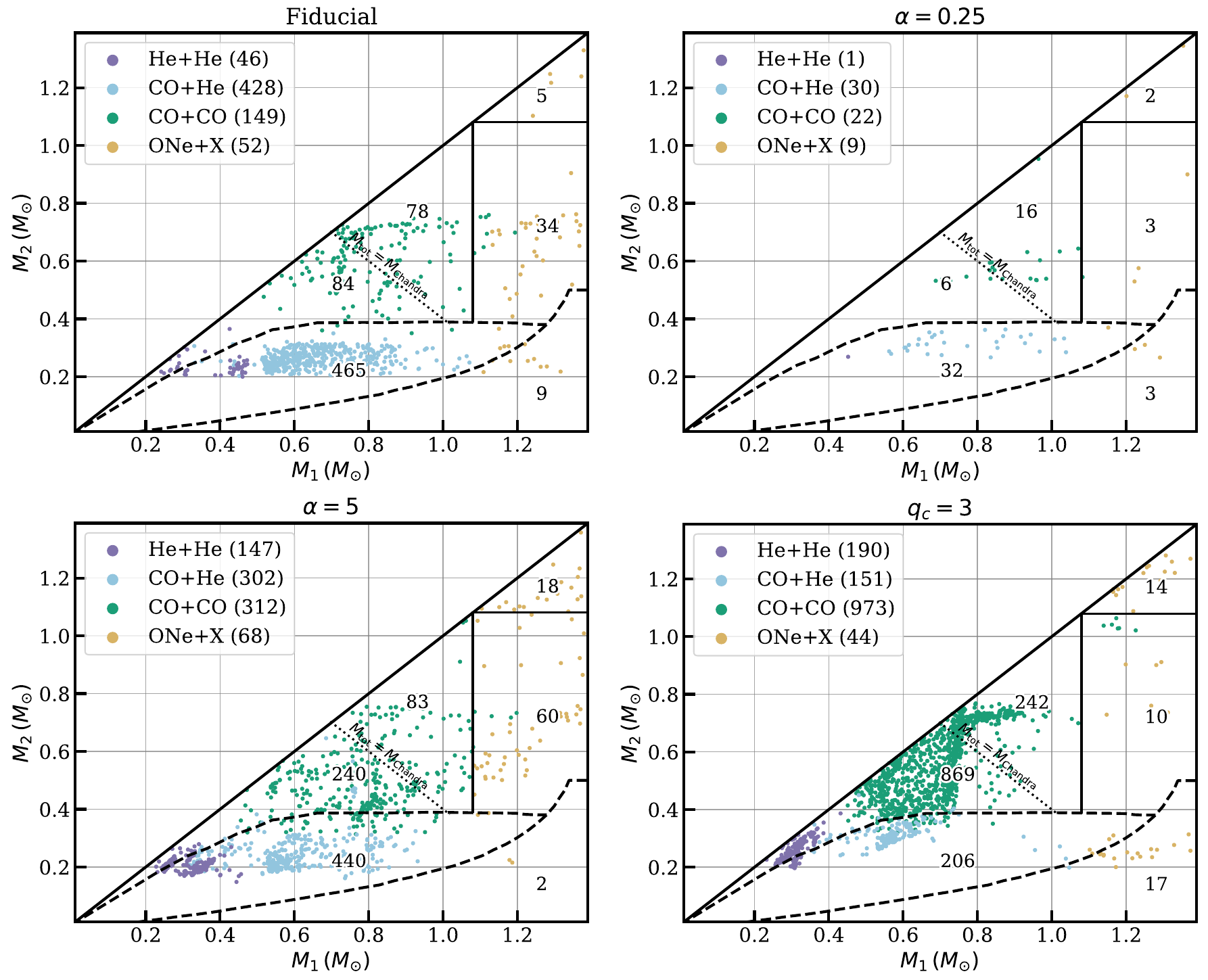}
    \caption{Same as Figure~\ref{fig:M1M2_all} but for only those white dwarf merger events occurring in the last $100\,$Myr and within $1\,$kpc of Earth.}
    \label{fig:M1M2_local}
\end{figure*}

\subsection{Milky Way populations of double white dwarf mergers}
We create simulated Milky-Way-like populations of double white dwarf mergers by assigning simulated double white dwarf binaries to star particles in the \texttt{m12i} galaxy of the Latte Suite of FIRE-2 galaxies \citep{Wetzel2016, Hopkins2018}. The \texttt{m12i} galaxy is approximately Milky-Way-like in terms of stellar and gas mass, size, and stellar morphology \citep[for detailed comparison to Galactic properties, see][]{Sanderson2020}, and has strong historical precedent for serving as a template for creating detailed synthetic binary populations related to white dwarf binaries and other sources \citep[e.g.,][]{Lamberts2018, Lamberts2019,Chawla2022,Thiele2023,DiCarlo2024,Tang2024,Delfavero2025,vanZeist2025}. For each star particle, we sample $N_{\rm{DWD}}$ double white dwarfs from the simulated population with the closest match in metallicity to the star particle, where 
\begin{equation}
    \label{eq:N_DWD}
    N_{\rm{DWD}}(Z) = n_{\rm{DWD,sim}}(Z) \frac{7000\,M_{\odot}}{m_{\rm{initial}}(Z)}.
\end{equation}
\noindent Here, the initial mass of each star particle is $7000\,M_{\odot}$ while $n_{\rm{DWD,sim}}$ is the number of double white dwarfs formed in a population initialized with mass $m_{\rm{initial}}$ for metallicity $Z$. Since $N_{\rm{DWD}}$ is not an integer by nature, we probabilistically add one double white dwarf based on the floating point value. Each sampled double white dwarf is assigned a ZAMS formation time based on the star particle age and a distance that is distributed with an Epanechnikov kernel centered on the present-day star particle position as described in \citet{Sanderson2020}.

Given a ZAMS formation time and simulated double white dwarf properties from \texttt{COSMIC}, we can evolve each simulated double white dwarf from formation to merger according to gravitational wave emission following \citet{Peters1964}.
The stability of mass transfer between white dwarfs is highly uncertain \citep[e.g.,][]{Marsh2004,Shen2015} so we do not separate white dwarf binaries that we expect to stably transfer mass from those that may merge. Instead, we holistically consider how the $M_1$-$M_2$ plane in Figure~\ref{fig:summary} is populated by our simulations. 

\section{Results}
\label{sec:results}

\subsection{Merger rate estimates}
\label{sec:rates}

In Figure~\ref{fig:M1M2_all} we plot secondary mass ($M_2$) versus primary mass ($M_1$) for all white dwarf binaries that come into Roche contact over the full history of each of our four Milky Way models (as labeled in each panel). Scatter points are colored based on the chemical composition of the white dwarf pair: purple denotes He+He, light blue denotes CO+He, dark green denotes CO+CO, and tan denotes pairs with an ONe primary and secondary of any composition (He, CO, or ONe). The legend for each panel notes the total number of each merger composition category occurring throughout that model's Galactic history. Each panel is divided into six regions corresponding to the six groups described in Figure~\ref{fig:summary}. The numbers marked within each region correspond to the merger rate for objects in that region averaged over the past $100\,$Myr. For example, in our ``Fiducial" model (top-left panel), we find a merger rate of $12.4\,\rm{Myr}^{-1}$ for region 6 (ONe+ONe mergers that we expect to ultimately collapse to a neutron star). We also list all merger rates for the different models in Table~\ref{table:rates}.

The total merger counts as well as the distributions in Figure~\ref{fig:M1M2_all} contain a number of distinct features illustrating how the binary evolution assumptions of each model imprint upon the white dwarf merger populations. We highlight here a few features.\footnote{Note that there is small amount of overlap between several of the populations in Figure~\ref{fig:M1M2_all}, especially the He+He and CO+He populations, which impacts the visual interpretation of the figure. We encourage a reader interested in exploring fine details of these populations to interact with the merger dataset itself (see Section~\ref{sec:data}).} First, we observe that larger $\alpha$ (increased common envelope efficiency) increases the total number of white dwarf mergers in strong agreement with \citet{YungelsonKuranov2017} (a factor of roughly $20$ for $\alpha5$ versus $\alpha0.25$). This is expected and consistent with previous studies; upon the onset of a common envelope, lower $\alpha$ corresponds to less efficient ejection of the envelope, which in general results in more failed common envelopes that lead to a stellar merger before a detached white dwarf binary can form.

The $q3$ model increases the critical mass ratio for onset of unstable mass transfer. Relative to our fiducial model, this enables higher mass donors to undergo stable mass transfer and avoid a common envelope. Comparing to the fiducial model, we see that higher $q_c$ increases only slightly the total number of white dwarf mergers ($7\times10^7$ to $10^8$). The relative numbers of different mass ratios of the white dwarf pairs are changed more substantially; the total number of mergers with near equal mass ratios (He+He, CO+CO, and ONe+ONe pairs) increases by a factor of 4.1, 4.5, and 1.8, respectively. However at $q_c=3$, the more asymmetric merger classes (He+CO, ONe+CO, ONe+He) are suppressed by a factor of 1.2, 1.6, and 1.5. At higher $q_c$, stable mass transfer is enabled for relatively massive donors (higher mass ratios). In these cases, a common envelope is avoided (typically during the initial phase of mass transfer), while stable mass transfer drives the system to equal component masses, and a near-equal mass white dwarf binary is ultimately formed. This is visible in the dearth of CO+CO and He+CO binaries with $q\lesssim 0.6$ in the $q3$ model.

We note that the enhancement of CO+CO white dwarfs (relative to He+CO white dwarfs) in our $q3$ is similar to predictions from population synthesis models performed with the \texttt{SeBa} code \citep[e.g.,][]{Nelemans2001a,Toonen2012,Korol2017}. This is likely due to \texttt{SeBa}'s application of the $\gamma$ prescription in the first common envelope phase, which produces binaries wider than the $\alpha\lambda$ common envelope prescription. This leads to fewer failed common envelopes during the second common envelope phase for CO+CO white dwarf binary progenitors. Although the increase in stable mass transfer events for the $q3$ model does not reproduce this effect in a one-to-one fashion, it qualitatively reduces the number of failed common envelopes for CO+CO white dwarf binaries due to wider separations after the first mass transfer phase. Also relevant to this discussion is \citet{Li2023}, which explored broadly how the criteria for mass transfer stability impact double white dwarf populations.

In all models, a ``cliff'' is visible just above $M_2\approx0.8\,M_\odot$. This arises when the common envelope that ensues when the secondary evolves off the main sequence strips the secondary's envelope before a sufficiently evolved core can form, inhibiting what would otherwise have formed a more massive white dwarf. The smaller number of CO+CO and CO+ONe sources in all models with $0.8 \lesssim M_2/M_\odot \lesssim 1.1$ are formed via mass ratio reversal during the initial phase of (stable) mass transfer. In this case the (now more massive secondary) builds a relatively massive core before its envelope is stripped via a common envelope when it ultimately evolves off the main sequence. A second cliff is visible in all models above $M_2 \gtrsim 0.4\,M_{\odot}$ for the CO+He and ONe+He binaries, arising from a roughly similar process where the second common envelope leaves behind the low-mass core of the secondary yet to begin helium burning.

In the $\alpha0.25$ model, the CO+CO and ONe+ONe mergers exhibit a pile up near $M_1=M_2$, a feature that is not present in the other models. This arises from roughly equal mass ZAMS binaries where, for lower $\alpha$, the second common envelope leaves a relatively compact CO+CO or ONe+ONe binary that for higher $\alpha$ would have been too wide to merge.

In Figure~\ref{fig:M1M2_local} we show only the white dwarf mergers occurring within $1\,$kpc and within the last $100\,$Myr. In practice, this is computed by counting all sources within a $1\,$kpc sphere centered around the Solar position (defined as $[G_x,G_y,G_z]=[8,0,0.015]\,$kpc) within the \texttt{m12i} FIRE-2 model. This local population is intended as a rough proxy for sources that may be observable today, however this of course depends on the details of the specific source formed. For example, the observable lifetime of young white dwarf formed by a sub-Chandrasekhar merger in Region 3 depends on its cooling time, while the lifetime of a neutron star formed via collapse in regions 5 or 6 depends on its spin-down time via magnetic dipole radiation. We discuss the observability of specific regions of this parameter space in Section~\ref{sec:outcomes}. Additionally, we caution again that \texttt{m12i} is a Milky Way-like analog, not a precise model. For example, \citet{Sanderson2020} notes that the present-day stellar density for the Solar neighborhood of \texttt{m12i} is roughly $30\%$ that of the observed value. In this case, the specific numbers quoted in Figure~\ref{fig:M1M2_local} and rates quoted in Table~\ref{table:rates} should be viewed with a factor of order unity uncertainty when attempting a one-to-one Milky Way extrapolation.

We note that despite the relatively small number of recent white dwarf mergers within $1\,$kpc, our different model assumptions lead to significant differences in both the relative rate of white dwarf merger type and mass. The trends observed in the entire Galactic population (Figure~\ref{fig:M1M2_all}) are broadly reflected in the population within $1\,$kpc shown in Figure~\ref{fig:M1M2_local}. This suggests that observations of local white dwarf merger products may provide key insights into the Galactic population of white dwarf binaries, including both merger products and inspiraling systems. The most obvious difference is the relative rate and mass ratio of He+CO and CO+CO binaries where models where common envelopes dominate the first phase of mass transfer produce more He+CO systems and models where stable mass transfer dominates produce more CO+CO systems. 

\begin{figure*}
    \centering
    \includegraphics[width=\linewidth]{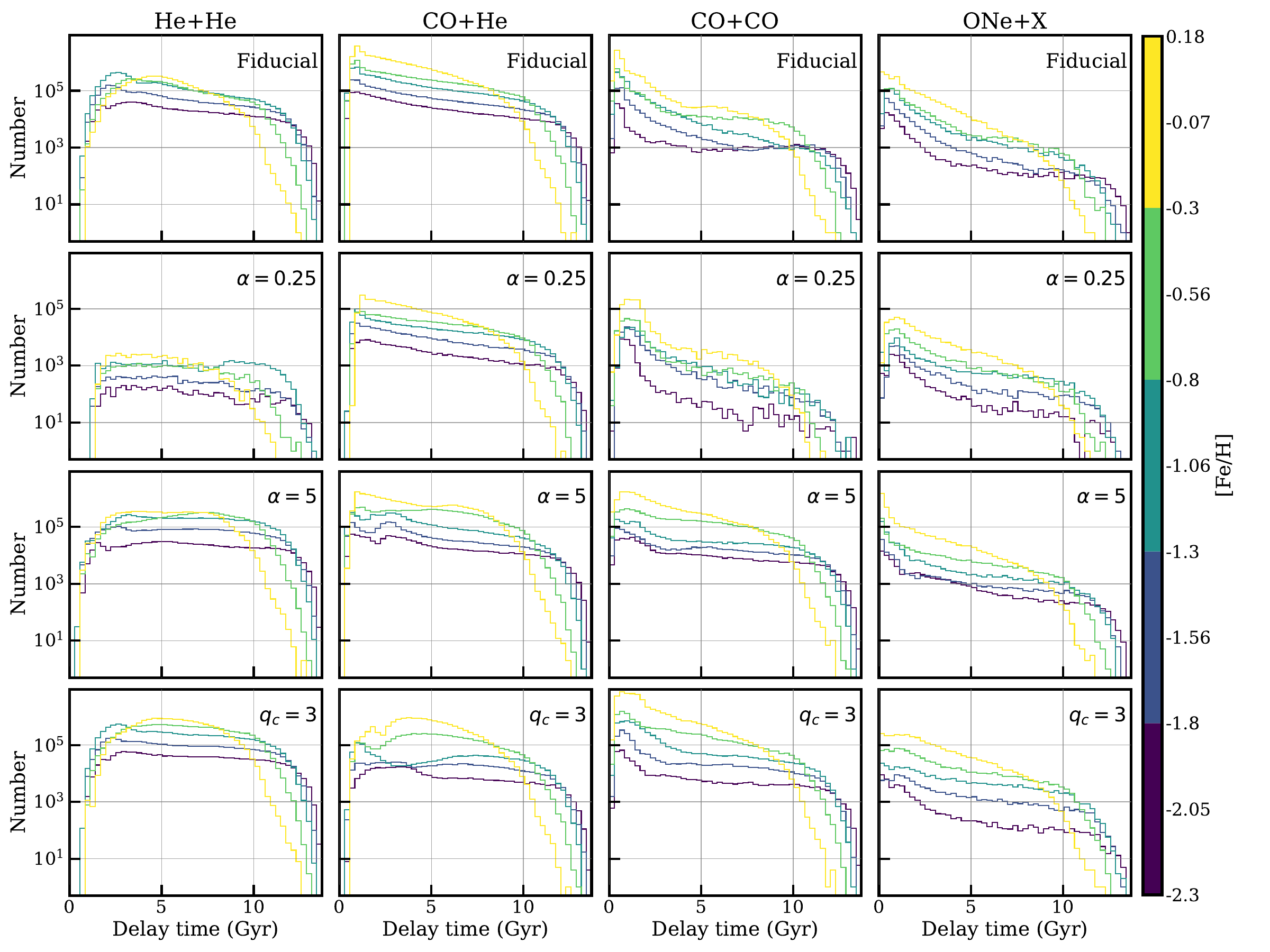}
    \caption{Delay time distributions (time from star formation to Roche contact) for all white dwarf pair combinations (columns from left to right) and Galactic model assumptions (rows from top to bottom). For each panel, we separate all white dwarf mergers into five metallicity bins centered on [Fe/H] =[-2.05, -1.56, -1.06, -0.56, -0.07].}
    \label{fig:DTDs}
\end{figure*}

\begin{figure*}
    \centering
    \includegraphics[width=\linewidth]{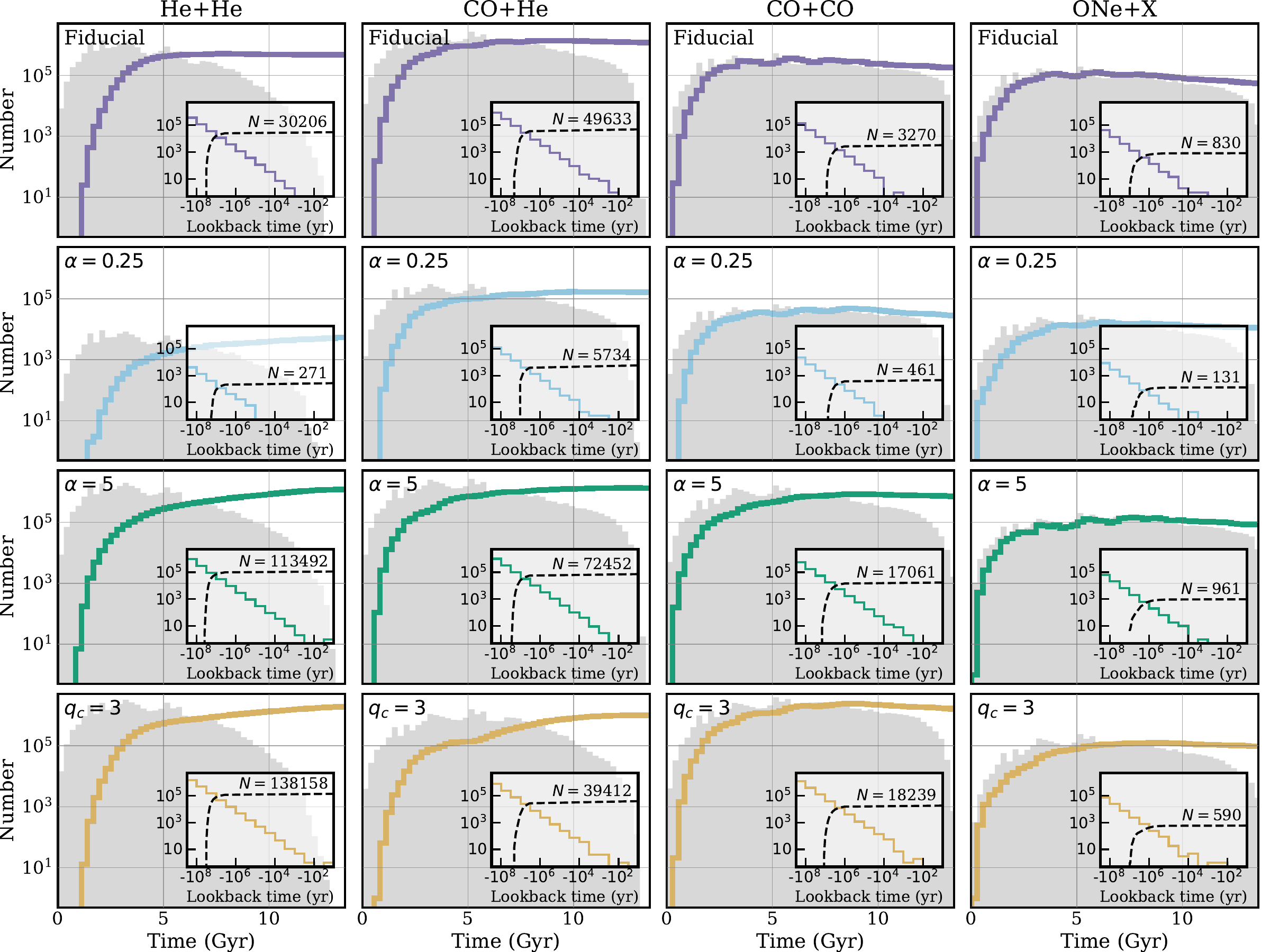}
    \caption{Colored curves show merger time distributions (in units of absolute Cosmic time) for all white dwarf pair combinations (columns) and binary evolution assumptions (rows). The solid gray histogram in each panel shows the distribution of star formation times for each white dwarf merger. In the insets, we zoom in on the most recent $100\,$Myr of Galactic history. Colored curves in the insets show the number of mergers at various lookback times. Dashed black curves show the cumulative number of white dwarf binaries formed with gravitational wave frequency $f_{\rm GW} > 10^{-3}\,$Hz (intended as a rough proxy for the number of resolvable LISA sources for each panel). As shown, panels featuring larger numbers of mergers typically correspond to larger number of LISA sources.}
    \label{fig:merger_times_WDtype}
\end{figure*}

\begin{figure*}
    \centering
    \includegraphics[width=\linewidth]{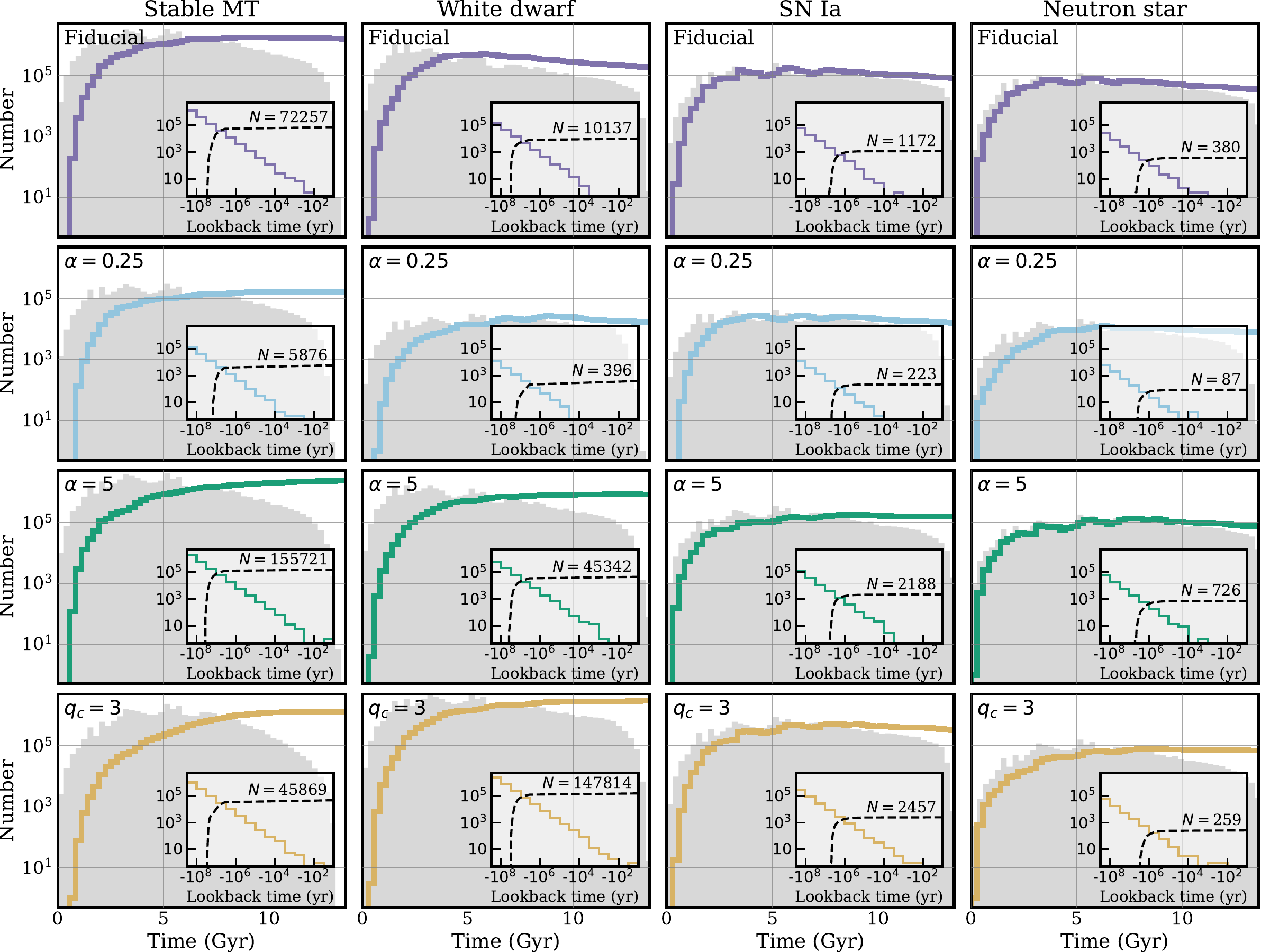}
    \caption{Same as Figure~\ref{fig:merger_times_WDtype} but grouping by merger outcome, rather than white dwarf pair combination. ``Stable MT'' corresponds to combined regions 1 and 2 in Figure~\ref{fig:summary}, ``white dwarf'' corresponds to region 3, ``SN Ia'' corresponds to region 4, and ``neutron star'' corresponds to combined regions 5 and 6.}
    \label{fig:merger_times_outcome}
\end{figure*}

\begin{deluxetable*}{l | p{2cm} p{2cm} p{2cm} p{2cm}}
\tablecaption{Recent (past $100\,$Myr) interaction rates for each Galactic model \label{table:rates}}
\tablehead{
\multicolumn{1}{l}{Outcome} &
\multicolumn{4}{c}{Rate ($\rm{yr}^{-1}$)} \\
\multicolumn{1}{l}{} &
\multicolumn{1}{l}{Fiducial} &
\multicolumn{1}{l}{$\alpha0.25$} &
\multicolumn{1}{l}{$\alpha5$} &
\multicolumn{1}{l}{$q3$}
}
\startdata
\textbf{All mergers} & $6.66\times10^{-3}$ & $7.39\times10^{-4}$ & $1.71\times10^{-2}$ & $1.35\times10^{-2}$ \\
\hline
He+He & $1.70\times10^{-3}$ & $1.90\times10^{-5}$ & $8.65\times10^{-3}$ & $6.94\times10^{-3}$ \\
He+CO & $4.13\times10^{-3}$ & $5.81\times10^{-4}$ & $5.61\times10^{-3}$ & $3.51\times10^{-3}$ \\
He+ONe & $6.56\times10^{-5}$ & $1.09\times10^{-5}$ & $2.58\times10^{-5}$ & $4.47\times10^{-5}$ \\
CO+CO & $6.37\times10^{-4}$ & $9.94\times10^{-5}$ & $2.53\times10^{-3}$ & $2.88\times10^{-3}$ \\
CO+ONe & $1.17\times10^{-4}$ & $2.34\times10^{-5}$ & $2.46\times10^{-4}$ & $7.14\times10^{-5}$ \\
ONe+ONe & $1.24\times10^{-5}$ & $4.93\times10^{-6}$ & $2.27\times10^{-5}$ & $2.18\times10^{-5}$ \\
\hline
\multicolumn{5}{c}{} \\
\hline
\textbf{Region 1 (Disk accretion)} & $5.53\times10^{-5}$ & $3.39\times10^{-6}$ & $3.29\times10^{-5}$ & $3.95\times10^{-5}$ \\
He+He & $0$ & $0$ & $0$ & $0$ \\
He+CO & $4.66\times10^{-6}$ & $0$ & $7.32\times10^{-6}$ & $1.66\times10^{-6}$ \\
He+ONe  & $4.62\times10^{-5}$ & $3.39\times10^{-6}$ & $2.12\times10^{-5}$ & $3.73\times10^{-5}$ \\
\hline
\textbf{Region 2 (Direct-impact accretion)} & $5.54\times10^{-3}$ & $5.90\times10^{-4}$ & $1.20\times10^{-2}$ & $4.20\times10^{-3}$ \\
He+He (SN .Ia) & $1.38\times10^{-3}$ & $9.48\times10^{-6}$ & $6.38\times10^{-3}$ & $7.26\times10^{-4}$ \\
He+CO (sub-Chandra Ia?) & $4.11\times10^{-3}$ & $5.74\times10^{-4}$ & $5.11\times10^{-3}$ & $3.27\times10^{-3}$ \\
\hline
\textbf{Region 3 (Sub-Chandra merger; single WD)} & $6.55\times10^{-4}$ & $5.75\times10^{-5}$ & $4.23\times10^{-3}$ & $8.55\times10^{-3}$ \\
He+He (sdB/O) & $3.16\times10^{-4}$ & $9.56\times10^{-6}$ & $2.27\times10^{-3}$ & $6.21\times10^{-3}$ \\
He+CO (sub-Chandra Ia?) & $1.88\times10^{-5}$ & $5.84\times10^{-6}$ & $4.96\times10^{-4}$ & $2.44\times10^{-4}$ \\
CO+CO & $3.20\times10^{-4}$ & $4.21\times10^{-5}$ & $1.47\times10^{-3}$ & $2.07\times10^{-3}$ \\
\hline
\textbf{Region 4 (Super-Chandra CO+CO)} & $2.91\times10^{-4}$ & $5.74\times10^{-5}$ & $5.31\times10^{-4}$ & $6.08\times10^{-4}$ \\
$q>0.8$ (Prompt C ignition \& SN Ia?) & $1.10\times10^{-4}$ & $1.39\times10^{-5}$ & $2.79\times10^{-4}$ & $5.36\times10^{-4}$ \\
$q<0.8$ (Off-center C ignition \& neutron star?) & $1.81\times10^{-4}$ & $4.35\times10^{-5}$ & $2.52\times10^{-4}$ & $7.28\times10^{-5}$ \\
\hline
\textbf{Region 5 (Super-Chandra ONe+CO; neutron star)} & $1.12\times10^{-4}$ & $2.34\times10^{-5}$ & $2.38\times10^{-4}$ & $7.10\times10^{-5}$ \\
\hline
\textbf{Region 6 (ONe+ONe; neutron star)}  & $1.24\times10^{-5}$ & $4.93\times10^{-6}$ & $2.27\times10^{-5}$ & $2.18\times10^{-5}$ \\
\hline
\hline
\enddata
\tablecomments{\footnotesize All rate values are averaged over past $100\,$Myr of Galactic evolution, and are expressed in units of $\rm{yr}^{-1}$.}
\end{deluxetable*}

\subsection{Delay time distributions}
\label{sec:delay_times}

In Figure~\ref{fig:DTDs} we show the delay time distribution (the time from zero-age main sequence, of the progenitor binary to Roche contact) for all white dwarf merger pairs. The four columns delineate the different four white dwarf pair combinations and the four rows delineate the four Galactic population models of Table~\ref{table:models}. Each panel is grouped into five metallicity bins over the range [Fe/H]$= [-2.3, 0.18]$ (as in \citet{Thiele2023}, we are intrinsically assuming [Fe/H] is approximately equal to $\log_{10}[Z/Z_\odot]$). 

Several trends are apparent: First, more massive white dwarfs (e.g., ONe+X compared to He+He) typically have shorter delay time distributions. For example in our Fiducial model, [$6\%$, $42\%$, $83\%$, $80\%$] of [He+He, CO+He, CO+CO, ONe+X] mergers have delay times less than $2\,$Gyr. This comes from two effects: (i) relatively massive progenitor stars have shorter stellar evolution lifetimes and (ii) once the white dwarf binary has formed, more massive white dwarfs have shorter gravitational wave inspiral times \citep[$t_{\rm insp}\propto~m^{-3}$ for $M_1=M_2=m$;][]{Peters1964}. Second, higher metallicity populations (e.g., compare yellow curves to purple curves) also typically have shorter delay times. Higher metallicity populations are formed relatively recently in the Galactic star formation history, meaning they have less time to reach Roche contact by present day compared to the low metallicity stars that formed further in the past. In this case, at higher metallicities, only the white dwarf binaries with shortest delay times reach Roche contact generally. There is a slight departure from this trend for the lowest mass white dwarfs (e.g. He+He), where the highest metallicity progenitors show a dearth of delay times below $1\,$Gyr. This is due to the more massive stellar progenitor filling it's Roche lobe while it is crossing the Hertzsprung Gap for high metallicities, rather than being on the main sequence for lower metallicities. In this case, a common envelope is initiated at higher metallicities and a stellar merger occurs during the second phase of mass transfer for the fiducial and $q3$ models. In the $\alpha=5$ variation, the second phase of mass transfer does not lead to a merger, while in the $\alpha=0.25$ variation, only the widest initial binaries survive such that the metallicity effect is negligible.

In Figure~\ref{fig:merger_times_WDtype}, we show as colored curves the \textit{merger times} of all white dwarf pairs. While the delay time denotes the time of merger relative to the star formation time of the progenitor binary, the merger time is measured relative to ``$t=0$'' of the Galaxy/Universe as a whole (i.e., also taking into account the star formation time of the progenitor binary). As in Figure~\ref{fig:DTDs}, columns delineate different white dwarf pair combinations while rows delineate the different Galactic models. Solid gray histograms show the distributions of star formation times for the progenitor binaries of each panel. The difference between the colored curves and solid gray histogram would be equivalent to the delay time distributions shown in Figure~\ref{fig:DTDs}.

The inset within each panel zooms in on the most recent snapshot of Galactic evolution, looking back from the present day to $100\,$Myr in the past. The colored curve in each inset shows the total number of mergers per (logarithmic) lookback time bin, illustrating the total number of sources of each type that have formed over a given lookback time. For example, for He+He white dwarf mergers in our Fiducial model (top-left panel of the figure), roughly 30 mergers occurred within the past $10^4\,$yr and roughly 1,700 mergers within the past $10^6\,$yr. The numeric value within each inset denotes the total number of binaries of each class with gravitational wave frequencies of at least $10^{-3}\,$Hz. This counts the binaries that have yet to merge and is intended as a rough proxy for the total number of sources of each class that would be individually resolvable by LISA \citep{Lamberts2019,LISA2023}. The dashed curve in each inset shows the \textit{cumulative} formation history of when this sample of binaries entered the $f_{\rm GW} > 10^{-3}\,$Hz frequency band. For example, from the top-left panel, we see that all $N=30,206$ He+He sources entered the millihertz frequency band within the past $4\times10^7\,$yr, while from the top right-panel, we see all $N=830$ ONe+X sources entered this band in the past $9\times10^6\,$yr. In this sense, the dashed curves are simply a measure of the inspiral time from $f_{\rm GW} = 10^{-3}\,$Hz to Roche contact for each class.

Figure~\ref{fig:merger_times_outcome} is similar to Figure~\ref{fig:merger_times_WDtype} but grouping instead based on the broadly-defined merger outcome instead of white dwarf combination, which we discuss further in Section~\ref{sec:outcomes}. Here ``Stable mass transfer'' combines all white dwarf interactions in Regions 1 and 2 of Figure~\ref{fig:summary}, ``white dwarf'' shows all mergers in Region 3, ``SN Ia'' all mergers in Region 4, and ``neutron star'' combines all mergers in Regions 5 and 6.

A key conclusion from both Figures~\ref{fig:merger_times_WDtype} and \ref{fig:merger_times_outcome} is that \textit{higher white dwarf merger rates correspond to larger numbers of LISA sources in the Milky Way today.} As a result, observations of these various white dwarf merger outcomes (described in detail in Section~\ref{sec:outcomes}) directly constrain the corresponding number of Galactic LISA source of that merger class.

\section{Merger outcomes and observable applications}
\label{sec:outcomes}

In this Section we discuss in further detail the various merger outcomes expected across our white dwarf parameter space, the relevant observable phases, and prospects for observational tests. We discuss each of the six main regions illustrated in Figure~\ref{fig:summary}. Our approach here is meant to be qualitative. More detailed follow-up studies of each merger class are necessary to ascertain in detail specific connections to the observed populations we highlight below.

\subsection{Stable mass transfer: AM CVn?}

For most asymmetric mass ratios, the onset of mass transfer will initially be dynamically stable and form an accretion disk \citep{Marsh2004}. Assuming that subsequent novae eruptions on the surface of the accretor enabled by burning of H or He do not drive the system to instability \citep{Shen2015}, these systems may undergo long-lived stable accretion. In general, binaries with mass ratios $q < 1$ will widen in response to mass transfer. However, for white dwarf binaries, additional effects also influence the orbital evolution including gravitational wave emission and tidal interactions between the two white dwarfs. Previous models predict that when an accretion disk forms, strong tidal coupling leads to efficient distribution of spin angular momentum back into the orbit, enabling the orbit to widen in opposition to gravitational wave inspiral \citep[e.g.,][]{Marsh2004}.

For mass ratios closer to unity, the accretor radius is sufficiently large relative to the orbital separation at Roche contact for the mass transfer to proceed via direct impact \citep{Nelemans2001b}. In the absence of an extended accretion disk, tides are expected to be less efficient at redistributing angular momentum back into the orbit, however in some cases, stable mass transfer may still be feasible \citep{Marsh2004,Gokhale2007,Kremer2015}. In cases where mass transfer is stable, the evolution is largely similar to the disk accretion case: as the system widens in response to mass transfer, its gravitational wave amplitude decreases. As the $\dot{M}$ values are generally higher at the onset of direct impact accretion relative to disk accretion, the lifetime spent in the direct impact phase is relatively short. Previous studies show the lifetime spent in the direct impact regime can vary from roughly $10^4-10^6\,$yr \citep{Kremer2017} compared to disk accretion timescales of roughly $10^8\,$yr or more, limited by the equation of state of the donor as it reaches extremely low mass \citep[e.g.,][]{Marsh2004}. We identify a formation rate for direct impact systems ranging from roughly $5\times10^{-4}\,\rm{yr}^{-1}$ ($\alpha0.25$) to roughly $10^{-2}\,\rm{yr}^{-1}$ ($\alpha5$). Adopting a direct impact lifetime of $10^4-10^6\,$yr, this implies anywhere from $10-10^4$ sources present in the Galaxy today, with potentially tens to hundreds of times more including systems in a disk configuration. Well over 100 AM CVn-type systems and related objects with roughly sub-hour orbital periods have now been identified observationally \citep[e.g.,][]{Solheim2010,Ramsay2018,vanRoestel2022,Green2025}.

The orbital evolution of these mass-transferring systems could produce a unique gravitational wave source that evolves to lower frequencies \citep{Kremer2017,Breivik2018}. The observable lifetime as a gravitational wave source depends on the depletion time of the donor (which reduces the binary's chirp mass and the strength of the gravitational wave signal), as well as the decrease in LISA's sensitivity at lower frequencies. In general, the systems easiest to observe in gravitational waves are high frequency sources that have recently started mass transfer \citep{Breivik2018}.

We raise one additional caveat concerning mergers involving at least one He white dwarf. He white dwarfs with masses sufficiently close to the expected He core mass at He flash (had the progenitor star not lost its envelope) may ultimately undergo He ignition and eventually become CO white dwarfs rather than proceeding directly to the He cooling track \citep[e.g.,][]{DCruz1996,Han2002}. The specific mass threshold for this process depends on many factors (stellar metallicity, progenitor mass, the details of the envelope stripping, the mass of any residual hydrogen layer leftover), but these previous studies generally quote roughly $0.45\,M_{\odot}$ as the minimum mass for He ignition. Across all models, roughly $3\%$ ($0.5\%$) of our He white dwarfs within He+He (CO+He) mergers have masses of $0.45\,M_{\odot}$ or more. Overall, we conclude consideration of this process would decrease slightly the rate of mergers involving He white dwarfs, but not a significant level. However, in the most extreme example in our $\alpha0.25$ model, $36\%$ of the He white dwarfs involved in He+He mergers have masses above this cut (see the pile up of He+He mergers at $M_1\approx 0.5\,M_{\odot}$ in the top-right panel of Figure~\ref{fig:M1M2_all}). In this specific case, this process may lead to an order unity correction to the He+He merger rates. These particular binaries would instead be categorized as He+CO mergers.

\vspace{1cm}

\subsection{Post-merger thermal evolution: Hot subdwarfs, R~Cor~Bor stars, and carbon giants}
\label{sec:RCrBr}

For mass ratios $q \gtrsim 2/3$, the mass transfer expected to be dynamically unstable \citep[e.g.,][]{Marsh2004} leading to a merger on dynamical timescales of order $10^2\,$s \citep[e.g.,][]{Benz1990,Guerrero2004,Yoon2007,LorenAguilar2009,Guillochon2010,Zhu2013,Dan2014,Tanikawa2019,Burmester2023}. Following the brief dynamical phase, the post-merger evolution is dictated by subsequent disk accretion of the disrupted material onto the accretor \citep[e.g.,][]{NomotoIben1985,Yoon2007,vanKerkwijk2010}. Modern studies argue that magnetohydrodynamic instabilities in the disk lead to a longer viscous phase lasting roughly $10^4-10^8\,$s \citep{Shen2012,Schwab2012}. Heating during the viscous phase transforms the initially lower mass white dwarf into a hot, slowly rotating, and radially extended envelope supported by thermal pressure. As thermal energy is radiated away, nuclear burning is triggered at the base of the envelope. The subsequent evolution, lifetime, and ultimate outcome of this thermal/nuclear phase depends on the composition and masses of the two white dwarfs. 

For the lowest mass mergers involving two He white dwarfs, off-center ignition of a He-burning shell leads to a series of He flashes that diffuse inwards on timescale $\sim10^6\,$yr \citep{Iben1990,SaioNomoto1998,Schwab2018}. These He flashes may cause the envelope to expand, and temporarily resemble an R~Cor~Bor star \citep{SaioJeffery2000,Justham2011}. Once the burning shell reaches the center, the merger remnant is compact ($\sim 0.1R_\odot$) and appears as a core He-burning star for roughly $10^8\,$yr until He-burning ceases. This scenario is believed to be a key channel for forming single sdB/sdO stars \citep[hot subdwarfs with hydrogen envelopes too small to support hydrogen shell burning; for review see][]{Heber2016}. In our models, we find a He+He white dwarf merger rate ranging from $2\times10^{-5}\,\rm{yr}^{-1}$ ($\alpha0.25$) to roughly $10^{-2}\,\rm{yr}^{-1}$ ($\alpha5$). In all models, at least $50\%$ (and for model $q3$, 90\%) of these lie within the direct impact accretion region, and may lead to stable mass transfer. Our predictions are roughly consistent with predictions in previous studies and with the observed sdOB population \citep{Nelemans2001a,Han2003,Dawson2024,Rodriguez-Segovia2025}. Following the hot subdwarf phase, the remnant ultimately cools to become a CO white dwarf \citep{Schwab2018}.

More massive ($0.8\,M_\odot \lesssim M_1+M_2 \lesssim1.4\,M_\odot$) He+CO white dwarf mergers will undergo He-shell burning and expand to become He giants with lifetime $\sim10^5\,$yr \citep{Paczynski1971,IbenTutukov1985}. This is expected to be a key channel for forming extreme He stars and R~Cor~Bor stars \citep{Webbink1984, IbenTutukov1985,Clayton1996,Schwab2019}. Our models predict a He+CO white dwarf merger rate ranging from $6\times10^{-4}\,\rm{yr}^{-1}$ ($\alpha0.25$) to $6\times10^{-3}\,\rm{yr}^{-1}$ ($\alpha5$). Assuming a lifetime of $\sim10^5\,$yr, this predicts order $10^2-10^3$ R~Cor~Bor stars in the Galaxy, comparable to the hundreds of sources currently known \citep[e.g.,][]{Tisserand2020}. Across all models, the majority of these (around 90\%) have mass ratios in our assumed direct impact regime. If, in fact, all of these direct impact sources avoid merger, our simulations may slightly underproduce the observed R~Cor~Bor population. Less than $1\%$ of our He+CO mergers have masses in the disk regime. Assuming a merger outcome, following He-shell burning, these remnants will ultimately cool to become (more massive) single CO white dwarfs.

The post-merger evolution of more massive mergers involving pairs of CO white dwarfs is more debated. For mass ratios near unity, the post-merger temperature and pressure may be sufficient to explosively ignite carbon in the center leading to a prompt explosion as a SN Ia \citep[e.g.,][see Section~\ref{sec:Ia}]{Pakmor2012,Dan2014}. For lower mass ratios, the donor is fully disrupted and the accretor may remain relatively cool, avoiding a prompt detonation. In this regime, subsequent viscous and thermal phases similar to the lower mass regime are expected \citep{Shen2012,Schwab2012}. An off-center carbon flame is ignited at the base of the envelope, which slowly propagates inward, transforming the CO core into ONeMg composition, lifting its degeneracy, and creating a luminous carbon giant with a lifetime roughly $10^4\,$yr \citep{Schwab2016,Schwab2021}. In our models, we find a CO+CO merger rate ranging from roughly $10^{-4}\,\rm{yr}^{-1}$ ($\alpha5$) to $3\times10^{-3}\,\rm{yr}^{-1}$ ($q3$), suggesting of order $1-30$ Galactic objects in the post-merger carbon giant phase today. \citet{Gvaramadze2019} report the detection of a hot, luminous object in a H-and-He-free nebula that roughly resembles the properties predicted for a CO+CO white dwarf merger remnant.

ONe+CO mergers are expected to similarly lead to a $\sim10^4\,$yr carbon giant phase. This case is differentiated from the CO+CO regime because the core remains degenerate. Instead, nuclear ash from carbon burning is deposited onto the degenerate ONe core until the mass becomes high enough for runaway electron captures to trigger collapse \citep[e.g.][]{SaioNomoto1985}. Our models predict an ONe+CO merger rate of roughly $7 \times 10^{-5}\,\rm{yr}^{-1}$ to $2 \times 10^{-4}\,\rm{yr}^{-1}$, implying order 1 post-merger carbon giants in the Milky Way from ONe+CO channel. 

\subsection{Formation of a new single white dwarf}

Following the thermal phase, the ultimate fate of a sub-Chandrasekhar\footnote{Note that uncertain mass loss during the luminous giant phase may in some cases remove enough mass for an initially super-Chandrasekhar merger to ultimately leave behind a massive white dwarf as well \citep[e.g.,][]{Schwab2021}.} white dwarf merger that avoids detonation (see Section~\ref{sec:Ia}) is a new, more massive, single white dwarf. In our models, the total rate of sub-Chandrasekhar mergers ranges from $7\times10^{-4}\,\rm{yr}^{-1}$ ($\alpha0.25$) to $1.6\times10^{-2}\,\rm{yr}^{-1}$ ($\alpha5$). Roughly 0.1\% and 80\% of these lie in regions 1 and 2, respectively, and thus may in fact undergo stable mass transfer and avoid merger. In this case, the formation rate of single white dwarfs would be reduced accordingly. Once the final white dwarf forms, the key timescale for observability is the cooling time.

A growing number of observations have associated specific white dwarf subpopulations with a white dwarf merger origin. One example is the subclass of isolated, rapidly spinning (sub-hour period), and highly magnetic (field strengths $\gtrsim 10^6\,$G) white dwarfs identified by SDSS \citep{Ferrario2015}. More recent surveys like the Zwicky Transient Facility have also linked new rapidly spinning, highly magnetic sources to a merger origin \citep{Caiazzo2021,Caiazzo2023}. Kinematic analysis of \textit{Gaia} data suggests roughly 20\% of all massive white dwarfs may be merger products \citep{Cheng2020}. The merger products are expected to have relatively high velocity dispersion because they are older than massive white dwarfs born via single star evolution. Additional work has explored a subclass of massive ($\approx 0.8-1.3\,M_{\odot}$) DQ white dwarfs with unique chemical and kinematic properties as merger products \citep{DunlapClemens2015,Coutu2019,KoesterKepler2019,Cheng2019}.

Note that the delay time distributions (time from ZAMS to merger) of Figure~\ref{fig:DTDs} feature significant numbers of sources with short delay times, roughly $1\,$Gyr or less. Indeed, the delay times for CO+CO binary mergers \textit{peak} at or below roughly $1\,$Gyr (although many of these are likely sufficiently massive to explode; Section~\ref{sec:Ia}). This indicates that long delay times are not necessarily a requirement for single white dwarfs formed via merger origin.

\subsection{Explosive outcomes and associated transients}
\label{sec:Ia}

Mergers of two CO white dwarfs with total mass in excess of the Chandrasekhar limit are a canonical scenario for SNe Ia \citep[e.g.,][]{IbenTutukov1984,Webbink1984}, with central carbon ignition triggering an explosion. More recent work has shown the details and pathways through which central carbon ignition and detonation occur are more complex. Hydrodynamics simulations show that ``violent mergers'' of two CO white dwarfs of roughly equal mass lead to central temperatures and densities sufficient for carbon ignition, triggering a prompt SN Ia explosion within a dynamical time of the merger \citep[e.g.,][]{Pakmor2012}. \citet{Dan2014} found that central carbon ignition may only occur for sufficiently massive pairs $M_1+M_2\gtrsim 2.1\,M_{\odot}$. For mass ratios further from unity (or lower masses), central carbon ignition does not occur at early times and a prompt explosion is avoided. In this non-destructive case, an off-center carbon flame will ultimately lead to merger-induced collapse of the remnant into a neutron star \citep{NomotoIben1985,SaioNomoto1985}, or a massive single white dwarf for sub-Chandrasekhar final masses. If the carbon flame lifts the degeneracy of the central remnant as it propagates inward, this process may proceed via the formation of a low-mass iron core \citep{Schwab2016,Schwab2021}, similar to the exposed, low-mass metal cores found in the progenitors of ultra-stripped supernovae \citep[e.g.,][]{Tauris2015}. The final collapse to a neutron star may itself be accompanied by a luminous transient distinct from a typical SN Ia. \citet{Brooks2017} showed that for expected explosion energies of roughly $10^{50}\,$ergs \citep{Kitaura2006,Dessart2006} and envelope masses of roughly $0.1\,M_\odot$, transient luminosities $\gtrsim 10^{43}\,\rm{erg\,s}^{-1}$ may be feasible, potentially similar to the class of rapidly evolving transients identified by \citet{Drout2014}.
In our models, we find that CO+CO mergers with $M_1+M_2>1.4\,M_\odot$ occur at rates ranging from $6\times10^{-5}\,\rm{yr}^{-1}$ ($\alpha0.25$) to $6\times10^{-4}\,\rm{yr}^{-1}$ ($q3$). In the $q3$ model, roughly 90\% of these have mass ratios $\gtrsim 0.8$ expected to be necessary for central carbon ignition. In the other three models, the fraction with $\gtrsim 0.8$ is slightly smaller, ranging from roughly 25\% to 50\%. For CO+CO mergers with $\lesssim 0.8$, collapse to a neutron star may be a more plausible outcome \citep[e.g.,][]{Schwab2021}.

If He is present \citep[either in the case of a He white dwarf donor, or CO white dwarf with a He surface layer; e.g.,][]{ShenMoore2014}, He detonations on the surface of the accretor may power bright explosions that may resemble faint SN Ia \citep[often referred to as ``.Ia'' events;][]{Bildsten2007}. In some cases, He detonation may drive a sufficiently strong shock into the CO core to trigger central carbon detonation, destroy the merger remnant entirely, and produce a SN Ia \citep{Pakmor2012}. This double-detonation scenario has been invoked as a popular pathway for sub-Chandrasekhar SNe Ia \citep[e.g.,][]{WoosleyWeaver1994,Fink2010}. Indeed, \citet{Shen2024} point out that because He surface layers are ubiquitous for CO white dwarfs, \textit{all} CO white dwarfs with mass $\lesssim 1.0\,M_\odot$ can support He detonations that may lead to a double detonation event. \citet{Shen2015} argued that even for low-mass ratio systems that may otherwise undergo stable mass transfer (Regions 1 and 2 of Figure~\ref{fig:summary}), novae eruptions from H or He burning may drive all interacting white dwarfs to dynamical instability, merger, and a possible double-detonation-powered SN Ia. As discussed in Section~\ref{sec:RCrBr}, the total He+CO white dwarf merger rate in our models ranges from $6\times10^{-4}\,\rm{yr}^{-1}$ ($\alpha0.25$) to $6\times10^{-3}\,\rm{yr}^{-1}$ ($\alpha5$), roughly 10 times higher than the CO+CO merger rate.

The SN Ia rate for the Milky Way is estimated to be roughly $10^{-3}-10^{-2}\,\rm{yr}^{-1}$, as inferred from observations of explosions in other nearby galaxies of similar mass \citep[e.g.,][]{Li2011,MaozMannucci2012,Liu2023,Munday2025}. Even in our most optimistic model $\alpha5$, our predicted super-Chandrasekhar merger rate (roughly $10^{-3}\,\rm{yr}^{-1}$) is lower than the inferred Galactic SN Ia rate, suggesting that sub-Chandrasekhar mergers must indeed contribute as is now widely accepted \citep[e.g.,][]{MaozMannucci2012,Maoz2014,YungelsonKuranov2017}. Assuming that all He+CO mergers also lead to detonation, we predict that our total rate becomes comparable to the inferred SN Ia rate. It is also useful to compare the observed Ia rate to the \textit{total} white dwarf merger rate, i.e., the extreme (and unrealistic) upper limit where \textit{all} interacting white dwarf binaries explode as a SN Ia. Our total rate ranges from roughly $7\times10^{-4}\,\rm{yr}^{-1}$ ($\alpha0.25$) to $1.7\times10^{-2}\,\rm{yr}^{-1}$ ($\alpha5$). Only our two most optimistic models, $\alpha5$ and $q3$, obviously overproduce the Galactic SN Ia rate.

\subsection{Collapse to a neutron star}

The most massive mergers that avoid explosion are expected to collapse to a neutron star. This includes a subset of CO+CO mergers, and likely all mergers with at least one ONe white dwarf. Conservation of angular momentum and magnetic flux during the final collapse is expected to yield neutron stars with small spin periods $\lesssim 10\,$ms and powerful magnetic fields $\gtrsim 10^{12}\,$G \citep[e.g.,][]{King2001,Schwab2021,Kremer2023}. Once formed, a neutron star evolving via magnetic dipole radiation will spin down, following tracks of roughly constant magnetic field strength

\begin{multline}
    \label{eq:B}
    B \approx \Bigg( \frac{3c^3I}{8\pi^2R_{\rm ns}^6} \Bigg)^{1/2} (P \dot{P} )^{1/2} \\ \approx 10^{12} \Bigg( \frac{P}{100\,\rm{ms}}\Bigg)^{1/2} \Bigg( \frac{\dot{P}}{10^{-14}\rm{s/s}}\Bigg)^{1/2}\,\rm{G}
\end{multline}
(here $I\approx 0.4 M R_{\rm ns}^2$ is the neutron star's moment of inertia) and crossing across lines of constant characteristic age

\begin{equation}
    \label{eq:tau}
    \tau \approx \frac{P}{2\dot{P}} \approx 10^5\, \Bigg( \frac{\it{P}}{50\,\rm{ms}}\Bigg)^2 \Bigg( \frac{\it{B}}{10^{12}\,\rm{G}}\Bigg)^{-2}\,\rm{yr}
\end{equation}
\citep[e.g.,][]{ShapiroTeukolsky1983}.

Eventually, pulsars spin down sufficiently to fall below the so-called ``death line'', an empirical boundary near $B/P^2 = 1.7\times10^{11}\,\rm{G\,s}^{-2}$ \citep[e.g.,][]{RudermanSutherland1975} below which they no longer are observable. For a given initial magnetic field, the characteristic spin-down time to reach this death line is roughly $\tau_{\rm sd} \approx 10^8 (B/10^{12}\,\rm{G})^{-1}\,\rm{yr}$, which is effectively the observable lifetime of a pulsar after formation.

In our models, the rate of super-Chandrasekhar mergers with at least one ONe white dwarf ranges from roughly $3\times10^{-5}\,\rm{yr}^{-1}$ ($\alpha0.25$) to $3\times10^{-4}\,\rm{yr}^{-1}$ ($\alpha5$). Presumably all of these will collapse to neutron stars. These rates may increase by a factor of roughly two if we also include super-Chandrasekhar CO+CO merges with $q < 0.8$ where central carbon detonation may be avoided. Assuming a characteristic lifetime $\tau_{\rm sd} \sim 10^8\,$yr, this equates to a population of roughly $10^3-10^5$ radio pulsars formed via white dwarf mergers in the Milky Way today. The observable sample is likely reduced by a factor of ten due to beaming for slow-spinning pulsars (spin periods $\gtrsim1\,{\rm s}$), but by less than 30\% for fast-spinning pulsars (spin periods $< 10\,{\rm ms}$) \citep[e.g.,][and their Figure~16]{Lorimer2008}. For reference, the rate of neutron stars formed via standard core collapse channel scales with the Galactic core-collapse SN rate, roughly $10^{-2}\,\rm{yr}^{-1}$, tens to hundreds of times higher than our predicted formation rate from white dwarf mergers.

Note that \citet{Shen2025} recently argued that the LP 40-365 class of O/Ne-rich hypervelocity stars \citep{Vennes2017,Raddi2018, ElBadry2023} are best explained as the bound remnants of ONe white dwarfs that underwent partial oxygen deflagrations, avoided runaway electron capture, and ultimately left behind a small unburnt remnant. In this case an ONe white dwarf that reaches the Chandrasekhar limit will not necessarily collapse in all cases. \citet{Shen2025} further advocate for an evolutionary outcome for ONe+CO mergers distinct from the canonical electron capture scenario \citep{Miyaji1980,Schwab2015}. In this picture, an ONe+CO merger results in an extended C-burning envelope that deposits ONe ash onto the surface of the ONe core. Eventually this ONe ash ignites and gives rise to successive phases of Ne, O, and Si burning the convert the initial ONe core into an Fe core, that ultimately undergoes an Fe core collapse and forms a neutron star, similar to the scenario described in \citet{Schwab2021} for pairs of massive CO white dwarfs. In this case, the canonical scenario where a collapse is facilitated by electron capture (see Region 5 of Figure~\ref{fig:summary}) may not be the most general case. The final outcome (formation of a neutron star) is expected to be unchanged, however the distinct evolutionary pathway may give rise to distinct observables during the pre-collapse and supernova phase.

\section{Discussion and Conclusions}
\label{sec:summary}

\subsection{Alternative formation channels}

We have focused exclusively on white dwarf mergers occurring via isolated binary evolution. A number of studies have explored alternatives. We briefly summarize two alternative here and reserve detailed comparisons across possible channels for future studies.

\textit{Globular clusters:} It is now well-established from direct observations of the white dwarf cooling sequence for a number of nearby systems that globular clusters host significant populations of white dwarfs \citep[e.g.,][]{Richer1995,Renzini1996,Cool1996,Hansen2013}. In these dense environments, dynamical encounters will inevitably lead to formation of white dwarf binary pairs and ultimately mergers \citep[e.g.,][]{Kremer2021}. Previous studies have argued that clusters may contribute appreciably to the local SN Ia rate \citep[e.g.,][]{SharaHurley2002}, however a SN Ia has yet be observed in association with a known globular cluster \citep[although for a candidate, see][]{Bregman2024}. Using N-body simulations, \citet{Kremer2021} showed that the white dwarf merger rate is expected to be particularly high in core-collapsed clusters. In these systems, the stellar black hole populations have been depleted, enabling massive white dwarfs to dominate the dynamics of the ultra-dense central regions. The total white dwarf merger rate in Galactic clusters is expected to be roughly $3\times10^{-7}\,\rm{yr}^{-1}$ \citep{Kremer2023}. The white dwarf mergers formed dynamically in globular clusters feature a significant bias toward the most massive mergers as a consequence of mass segregation; 90\% of white dwarf mergers in clusters have a total mass in excess of the Chandrasekhar limit \citep{Kremer2021}. In this case, these are strong candidates for undergoing collapse to neutron stars and, as a result, have been connected to a subpopulation of apparently young radio pulsars in Galactic globular clusters \citep{Kremer2023} and also to a recent repeating fast radio burst in a globular cluster in M81 \citep{Kristen2022,Kremer2021_frb,Lu2022,Rao2025}.

Assuming that the ratio of total stellar mass of the full Galactic globular cluster system to the Milky Way is roughly $10^{-4}$, this implies that globular clusters feature an enhancement of roughly $30-100$ in the formation rate per stellar mass of neutron stars via collapsing white dwarf mergers. The upper limit in this range is roughly comparable to the globular cluster enhancement of roughly $200$ for low-mass X-ray binaries and millisecond pulsars \citep[e.g.,][]{Clark1975,Katz1975,Pooley2003,Bahramian2013,Kremer2026}, and is slightly larger than the upper limit of $50$ placed on the enhancement of SN Ia in globular clusters based on null detections \citep{VossNelemans2012,WashabaughBregman2013,Bregman2024}.

\textit{Stellar triples or higher multiples:} Roughly a third to half of main sequence stars with masses of roughly $2-8\,M_{\odot}$ (the expected progenitors of white dwarfs) are members of stellar triples \citep[e.g.,][]{Raghavan2010,Tokovinin2014,MoeDiStefano2017,Offner2023}. The vast majority of these are hierarchical triples \citep[e.g.,][]{Shariat2025} subject to the eccentric Lidov-Kozai mechanism \citep{Kozai1962,Lidov1962,Naoz2016}. A number of studies have demonstrated that this effect may enhance the rate of white dwarf mergers in these hierarchical triples systems \citep{THompson2011,AntogniniThompson2014,Fang2018,Toonen2018,Sheriat2026}. For example, \citet{Shariat+2026} demonstrated that white dwarf merger rates in hierarchical triples can match or exceed those from isolated binary evolution in the local Universe, despite the relatively high binary fraction among low-mass stars. \citet{Rajamuthukumar2025} recently performed simulations combining stellar evolution with secular effects and specifically showed that white dwarfs in triples are an important class of sources for LISA. Future work may perform a direct comparison of white dwarf mergers among all relevant channels to explore the interplay of binary evolution and dynamical processes in the full merger population.

\subsection{SN Ia and galaxy feedback}

Supernovae play an important role in chemical and mechanical feedback processes in galaxies 
\citep[e.g.,][]{Hopkins2014,SomervilleDave2015,NaabOstriker2017,Kobayashi2020}. Indeed, the FIRE-2 simulations used in this study as a proxy for a Milky Way analog incorporate feedback from both Type II and Type Ia SNe, using simple fits to mimic stellar evolution and associated yields \citep[for implementation details, see Appendix A of][]{Hopkins2018}. These feedback processes ultimately play an important role in subsequent star formation \citep[e.g.,][]{OstrikerShetty2011,Hopkins2014}, and therefore on the subsequent SN Ia rate. In this case, our current procedure of stitching together FIRE simulations and \texttt{COSMIC} population synthesis models is not self-consistent; changes to the SN Ia delay-time distribution, normalization, energetics, yields, and/or spatial distribution implied by a specific binary evolution model/white dwarf merger model could, in principle, alter the chemical enrichment and star-formation history of the host galaxy itself. Ideally, a SN Ia feedback model would be implemented ``on-the-fly'' with a galaxy simulation, motivated by results presented here for white dwarf mergers. We reserve such study for future work.

\subsection{Summary}

We have computed the Milky Way white dwarf merger history for several different assumptions pertaining to Roche lobe overflow interactions. 
We remain agnostic to the specific outcome resulting from the onset of mass transfer for double white dwarf binaries (e.g., whether mass transfer is dynamically stable or unstable) and instead summarize all possibilities discussed in the literature (see Figure~\ref{fig:summary}).
We connect the recent white dwarf merger rate to various observed sources including AM CVn binaries, R~Cor~Bor stars, single white dwarfs from low-mass mergers, SN Ia, and neutron stars/pulsars born via collapse. The relative formation rates of these sources vary across our different Galactic models by a factor of ten or more, demonstrating the sensitivity of these mergers to binary evolution prescriptions.

\citet{Thiele2023} used these same Galactic models to predict the white dwarf population detectable by LISA both as individually resolved sources and as foreground noise. The white dwarf merger products studied here represent the final outcome of millihertz LISA sources. Indeed, a key result of this study (see Figures~\ref{fig:merger_times_WDtype} and \ref{fig:merger_times_outcome}) is that the recent (e.g., within the past $100\,$Myr) white dwarf merger rate correlates directly with the number of LISA sources present in the Milky Way today. Therefore, observations of the outcomes of white dwarf mergers offer a powerful complementary tool for constraining estimates for Galactic LISA sources.

In a complementary example of this exercise, \citet{Korol2022} recently performed an empirical estimate of the Galactic white dwarf LISA sample using data from the Sloan Digital Sky Survey (SDSS) and the Supernova Ia Progenitor surveY (SPY). This study found that these observationally-driven estimates predict up to five times more individually detectable white dwarf binaries compared to typical binary population synthesis studies. This underlines the importance of using the many observational diagnostics outlined here as tools for better constraining population synthesis models and ultimately, predictions for the LISA population. To facilitate future similar studies in this arena, all merger data from these Galactic models are now available for download on Zenodo, along with corresponding analysis tools on Github.

\section{Data Availability}
\label{sec:data}

All white dwarf merger data produced in this study are available for download via Zenodo at DOI: \href{https://zenodo.org/records/19615369}{10.5281/zenodo.19615369}. All scripts and supplementary data required to reproduce our figures are available in Zenodo at DOI: \href{https://doi.org/10.5281/zenodo.21381908}{10.5281/zenodo.21381908} and on \href{https://github.com/kylekremer23/WDmergers_COSMIC}{GitHub}.

\acknowledgments

We thank the anonymous referee for a constructive report. K.K., K.B., and C.S.Y. acknowledge support from NASA LISA Preparatory Science Program Grant No. 80NSSC26K0339. K.B. acknowledges support from the Falco-DeBenedetti Early Career Professorship and NASA LISA Preparatory Science Program Grant No. 80NSSC24K0361. C.S.Y. acknowledges support from the Alfred P. Sloan Foundation. This research was supported in part by grant NSF PHY-2309135 to the Kavli Institute for Theoretical Physics (KITP).

\bibliographystyle{aasjournal}
\bibliography{mybib.bib}

@ARTICLE{Belczynski2008,
       author = {{Belczynski}, Krzysztof and {Kalogera}, Vassiliki and {Rasio}, Frederic A. and {Taam}, Ronald E. and {Zezas}, Andreas and {Bulik}, Tomasz and {Maccarone}, Thomas J. and {Ivanova}, Natalia},
        title = "{Compact Object Modeling with the StarTrack Population Synthesis Code}",
      journal = {\apjs},
         year = 2008,
        month = jan,
       volume = {174},
       number = {1},
        pages = {223-260},
          doi = {10.1086/521026},
archivePrefix = {arXiv},
       eprint = {astro-ph/0511811},
 primaryClass = {astro-ph},
       adsurl = {https://ui.adsabs.harvard.edu/abs/2008ApJS..174..223B}
}

@ARTICLE{YungelsonKuranov2017,
       author = {{Yungelson}, L.~R. and {Kuranov}, A.~G.},
        title = "{Merging white dwarfs and Type Ia supernovae}",
      journal = {\mnras},
         year = 2017,
        month = jan,
       volume = {464},
       number = {2},
        pages = {1607-1632},
          doi = {10.1093/mnras/stw2432},
       adsurl = {https://ui.adsabs.harvard.edu/abs/2017MNRAS.464.1607Y}
}

@ARTICLE{Brown2020,
       author = {{Brown}, Warren R. and {Kilic}, Mukremin and {Kosakowski}, Alekzander and {Andrews}, Jeff J. and {Heinke}, Craig O. and {Ag{\"u}eros}, Marcel A. and {Camilo}, Fernando and {Gianninas}, A. and {Hermes}, J.~J. and {Kenyon}, Scott J.},
        title = "{The ELM Survey. VIII. Ninety-eight Double White Dwarf Binaries}",
      journal = {\apj},
         year = 2020,
        month = jan,
       volume = {889},
       number = {1},
          eid = {49},
        pages = {49},
          doi = {10.3847/1538-4357/ab63cd},
archivePrefix = {arXiv},
       eprint = {2002.00064},
 primaryClass = {astro-ph.SR},
       adsurl = {https://ui.adsabs.harvard.edu/abs/2020ApJ...889...49B}
}

@ARTICLE{Brown2016,
       author = {{Brown}, Warren R. and {Kilic}, Mukremin and {Kenyon}, Scott J. and {Gianninas}, A.},
        title = "{Most Double Degenerate Low-mass White Dwarf Binaries Merge}",
      journal = {\apj},
         year = 2016,
        month = jun,
       volume = {824},
       number = {1},
          eid = {46},
        pages = {46},
          doi = {10.3847/0004-637X/824/1/46},
archivePrefix = {arXiv},
       eprint = {1604.04269},
 primaryClass = {astro-ph.SR},
       adsurl = {https://ui.adsabs.harvard.edu/abs/2016ApJ...824...46B}
}

@ARTICLE{Toonen2012,
       author = {{Toonen}, S. and {Nelemans}, G. and {Portegies Zwart}, S.},
        title = "{Supernova Type Ia progenitors from merging double white dwarfs. Using a new population synthesis model}",
      journal = {\aap},
         year = 2012,
        month = oct,
       volume = {546},
          eid = {A70},
        pages = {A70},
          doi = {10.1051/0004-6361/201218966},
archivePrefix = {arXiv},
       eprint = {1208.6446},
 primaryClass = {astro-ph.HE},
       adsurl = {https://ui.adsabs.harvard.edu/abs/2012A&A...546A..70T}
}

@ARTICLE{Korol2022,
       author = {{Korol}, Valeriya and {Hallakoun}, Na'ama and {Toonen}, Silvia and {Karnesis}, Nikolaos},
        title = "{Observationally driven Galactic double white dwarf population for LISA}",
      journal = {\mnras},
         year = 2022,
        month = apr,
       volume = {511},
       number = {4},
        pages = {5936-5947},
          doi = {10.1093/mnras/stac415},
archivePrefix = {arXiv},
       eprint = {2109.10972},
 primaryClass = {astro-ph.HE},
       adsurl = {https://ui.adsabs.harvard.edu/abs/2022MNRAS.511.5936K}
}

@ARTICLE{Sanderson2020,
       author = {{Sanderson}, Robyn E. and {Wetzel}, Andrew and {Loebman}, Sarah and {Sharma}, Sanjib and {Hopkins}, Philip F. and {Garrison-Kimmel}, Shea and {Faucher-Gigu{\`e}re}, Claude-Andr{\'e} and {Kere{\v{s}}}, Du{\v{s}}an and {Quataert}, Eliot},
        title = "{Synthetic Gaia Surveys from the FIRE Cosmological Simulations of Milky Way-mass Galaxies}",
      journal = {\apjs},
         year = 2020,
        month = jan,
       volume = {246},
       number = {1},
          eid = {6},
        pages = {6},
          doi = {10.3847/1538-4365/ab5b9d},
archivePrefix = {arXiv},
       eprint = {1806.10564},
 primaryClass = {astro-ph.GA},
       adsurl = {https://ui.adsabs.harvard.edu/abs/2020ApJS..246....6S}
}

@ARTICLE{Shen2012,
       author = {{Shen}, Ken J. and {Bildsten}, Lars and {Kasen}, Daniel and {Quataert}, Eliot},
        title = "{The Long-term Evolution of Double White Dwarf Mergers}",
      journal = {\apj},
         year = 2012,
        month = mar,
       volume = {748},
       number = {1},
          eid = {35},
        pages = {35},
          doi = {10.1088/0004-637X/748/1/35},
archivePrefix = {arXiv},
       eprint = {1108.4036},
 primaryClass = {astro-ph.HE},
       adsurl = {https://ui.adsabs.harvard.edu/abs/2012ApJ...748...35S}
}

@ARTICLE{Schwab2016,
       author = {{Schwab}, Josiah and {Quataert}, Eliot and {Kasen}, Daniel},
        title = "{The evolution and fate of super-Chandrasekhar mass white dwarf merger remnants}",
      journal = {\mnras},
         year = 2016,
        month = dec,
       volume = {463},
       number = {4},
        pages = {3461-3475},
          doi = {10.1093/mnras/stw2249},
archivePrefix = {arXiv},
       eprint = {1606.02300},
 primaryClass = {astro-ph.SR},
       adsurl = {https://ui.adsabs.harvard.edu/abs/2016MNRAS.463.3461S}
}

@ARTICLE{Schwab2012,
       author = {{Schwab}, Josiah and {Shen}, Ken J. and {Quataert}, Eliot and {Dan}, Marius and {Rosswog}, Stephan},
        title = "{The viscous evolution of white dwarf merger remnants}",
      journal = {\mnras},
         year = 2012,
        month = nov,
       volume = {427},
       number = {1},
        pages = {190-203},
          doi = {10.1111/j.1365-2966.2012.21993.x},
archivePrefix = {arXiv},
       eprint = {1207.0512},
 primaryClass = {astro-ph.HE},
       adsurl = {https://ui.adsabs.harvard.edu/abs/2012MNRAS.427..190S}
}

@ARTICLE{Dan2014,
       author = {{Dan}, Marius and {Rosswog}, Stephan and {Br{\"u}ggen}, Marcus and {Podsiadlowski}, Philipp},
        title = "{The structure and fate of white dwarf merger remnants}",
      journal = {\mnras},
         year = 2014,
        month = feb,
       volume = {438},
       number = {1},
        pages = {14-34},
          doi = {10.1093/mnras/stt1766},
archivePrefix = {arXiv},
       eprint = {1308.1667},
 primaryClass = {astro-ph.HE},
       adsurl = {https://ui.adsabs.harvard.edu/abs/2014MNRAS.438...14D}
}

@ARTICLE{Dan2011,
       author = {{Dan}, Marius and {Rosswog}, Stephan and {Guillochon}, James and {Ramirez-Ruiz}, Enrico},
        title = "{Prelude to A Double Degenerate Merger: The Onset of Mass Transfer and Its Impact on Gravitational Waves and Surface Detonations}",
      journal = {\apj},
         year = 2011,
        month = aug,
       volume = {737},
       number = {2},
          eid = {89},
        pages = {89},
          doi = {10.1088/0004-637X/737/2/89},
archivePrefix = {arXiv},
       eprint = {1101.5132},
 primaryClass = {astro-ph.HE},
       adsurl = {https://ui.adsabs.harvard.edu/abs/2011ApJ...737...89D}
}

@ARTICLE{Nelemans2001a,
       author = {{Nelemans}, G. and {Yungelson}, L.~R. and {Portegies Zwart}, S.~F. and {Verbunt}, F.},
        title = "{Population synthesis for double white dwarfs . I. Close detached systems}",
      journal = {\aap},
         year = 2001,
        month = jan,
       volume = {365},
        pages = {491-507},
          doi = {10.1051/0004-6361:20000147},
archivePrefix = {arXiv},
       eprint = {astro-ph/0010457},
 primaryClass = {astro-ph},
       adsurl = {https://ui.adsabs.harvard.edu/abs/2001A&A...365..491N}
}

@ARTICLE{Shen2018,
       author = {{Shen}, Ken J. and {Boubert}, Douglas and {G{\"a}nsicke}, Boris T. and {Jha}, Saurabh W. and {Andrews}, Jennifer E. and {Chomiuk}, Laura and {Foley}, Ryan J. and {Fraser}, Morgan and {Gromadzki}, Mariusz and {Guillochon}, James and {Kotze}, Marissa M. and {Maguire}, Kate and {Siebert}, Matthew R. and {Smith}, Nathan and {Strader}, Jay and {Badenes}, Carles and {Kerzendorf}, Wolfgang E. and {Koester}, Detlev and {Kromer}, Markus and {Miles}, Broxton and {Pakmor}, R{\"u}diger and {Schwab}, Josiah and {Toloza}, Odette and {Toonen}, Silvia and {Townsley}, Dean M. and {Williams}, Brian J.},
        title = "{Three Hypervelocity White Dwarfs in Gaia DR2: Evidence for Dynamically Driven Double-degenerate Double-detonation Type Ia Supernovae}",
      journal = {\apj},
         year = 2018,
        month = sep,
       volume = {865},
       number = {1},
          eid = {15},
        pages = {15},
          doi = {10.3847/1538-4357/aad55b},
archivePrefix = {arXiv},
       eprint = {1804.11163},
 primaryClass = {astro-ph.SR},
       adsurl = {https://ui.adsabs.harvard.edu/abs/2018ApJ...865...15S}
}

@ARTICLE{ShenBildsten2009,
       author = {{Shen}, Ken J. and {Bildsten}, Lars},
        title = "{Unstable Helium Shell Burning on Accreting White Dwarfs}",
      journal = {\apj},
         year = 2009,
        month = jul,
       volume = {699},
       number = {2},
        pages = {1365-1373},
          doi = {10.1088/0004-637X/699/2/1365},
archivePrefix = {arXiv},
       eprint = {0903.0654},
 primaryClass = {astro-ph.HE},
       adsurl = {https://ui.adsabs.harvard.edu/abs/2009ApJ...699.1365S}
}

@ARTICLE{Bildsten2007,
       author = {{Bildsten}, Lars and {Shen}, Ken J. and {Weinberg}, Nevin N. and {Nelemans}, Gijs},
        title = "{Faint Thermonuclear Supernovae from AM Canum Venaticorum Binaries}",
      journal = {\apjl},
         year = 2007,
        month = jun,
       volume = {662},
       number = {2},
        pages = {L95-L98},
          doi = {10.1086/519489},
archivePrefix = {arXiv},
       eprint = {astro-ph/0703578},
 primaryClass = {astro-ph},
       adsurl = {https://ui.adsabs.harvard.edu/abs/2007ApJ...662L..95B}
}

@ARTICLE{ShenBildsten2014,
       author = {{Shen}, Ken J. and {Bildsten}, Lars},
        title = "{The Ignition of Carbon Detonations via Converging Shock Waves in White Dwarfs}",
      journal = {\apj},
         year = 2014,
        month = apr,
       volume = {785},
       number = {1},
          eid = {61},
        pages = {61},
          doi = {10.1088/0004-637X/785/1/61},
archivePrefix = {arXiv},
       eprint = {1305.6925},
 primaryClass = {astro-ph.HE},
       adsurl = {https://ui.adsabs.harvard.edu/abs/2014ApJ...785...61S}
}

@ARTICLE{Livne1990,
       author = {{Livne}, Eli},
        title = "{Successive Detonations in Accreting White Dwarfs as an Alternative Mechanism for Type I Supernovae}",
      journal = {\apjl},
         year = 1990,
        month = may,
       volume = {354},
        pages = {L53},
          doi = {10.1086/185721},
       adsurl = {https://ui.adsabs.harvard.edu/abs/1990ApJ...354L..53L}
}

@ARTICLE{Taam1980,
       author = {{Taam}, R.~E.},
        title = "{The long-term evolution of accreting carbon white dwarfs}",
      journal = {\apj},
         year = 1980,
        month = dec,
       volume = {242},
        pages = {749-755},
          doi = {10.1086/158509},
       adsurl = {https://ui.adsabs.harvard.edu/abs/1980ApJ...242..749T}
}

@ARTICLE{Carter2013,
       author = {{Carter}, P.~J. and {Marsh}, T.~R. and {Steeghs}, D. and {Groot}, P.~J. and {Nelemans}, G. and {Levitan}, D. and {Rau}, A. and {Copperwheat}, C.~M. and {Kupfer}, T. and {Roelofs}, G.~H.~A.},
        title = "{A search for the hidden population of AM CVn binaries in the Sloan Digital Sky Survey}",
      journal = {\mnras},
         year = 2013,
        month = mar,
       volume = {429},
       number = {3},
        pages = {2143-2160},
          doi = {10.1093/mnras/sts485},
archivePrefix = {arXiv},
       eprint = {1211.6439},
 primaryClass = {astro-ph.SR},
       adsurl = {https://ui.adsabs.harvard.edu/abs/2013MNRAS.429.2143C}
}

@ARTICLE{Podsiadlowski2003,
       author = {{Podsiadlowski}, Ph. and {Han}, Z. and {Rappaport}, S.},
        title = "{Cataclysmic variables with evolved secondaries and the progenitors of AM CVn stars}",
      journal = {\mnras},
         year = 2003,
        month = apr,
       volume = {340},
       number = {4},
        pages = {1214-1228},
          doi = {10.1046/j.1365-8711.2003.06380.x},
       adsurl = {https://ui.adsabs.harvard.edu/abs/2003MNRAS.340.1214P}
}

@ARTICLE{IbenTutukov1987,
       author = {{Iben}, Jr., Icko and {Tutukov}, Alexander V.},
        title = "{Evolutionary Scenarios for Intermediate-Mass Stars in Close Binaries}",
      journal = {\apj},
         year = 1987,
        month = feb,
       volume = {313},
        pages = {727},
          doi = {10.1086/165011},
       adsurl = {https://ui.adsabs.harvard.edu/abs/1987ApJ...313..727I}
}

@ARTICLE{Savonije1986,
       author = {{Savonije}, G.~J. and {de Kool}, M. and {van den Heuvel}, E.~P.~J.},
        title = "{The minimum orbital period for ultra-compact binaries with the helium burning secondaries.}",
      journal = {\aap},
         year = 1986,
        month = jan,
       volume = {155},
        pages = {51-57},
       adsurl = {https://ui.adsabs.harvard.edu/abs/1986A&A...155...51S}
}

@ARTICLE{Kremer2015,
       author = {{Kremer}, Kyle and {Sepinsky}, Jeremy and {Kalogera}, Vassiliki},
        title = "{Long-term Evolution of Double White Dwarf Binaries Accreting through Direct Impact}",
      journal = {\apj},
         year = 2015,
        month = jun,
       volume = {806},
       number = {1},
          eid = {76},
        pages = {76},
          doi = {10.1088/0004-637X/806/1/76},
archivePrefix = {arXiv},
       eprint = {1502.06147},
 primaryClass = {astro-ph.SR},
       adsurl = {https://ui.adsabs.harvard.edu/abs/2015ApJ...806...76K}
}

@ARTICLE{Gokhale2007,
       author = {{Gokhale}, Vayujeet and {Peng}, Xiao Meng and {Frank}, Juhan},
        title = "{Evolution of Close White Dwarf Binaries}",
      journal = {\apj},
         year = 2007,
        month = feb,
       volume = {655},
       number = {2},
        pages = {1010-1024},
          doi = {10.1086/510119},
archivePrefix = {arXiv},
       eprint = {astro-ph/0610919},
 primaryClass = {astro-ph},
       adsurl = {https://ui.adsabs.harvard.edu/abs/2007ApJ...655.1010G}
}

@ARTICLE{ElBadry2023,
       author = {{El-Badry}, Kareem and {Shen}, Ken J. and {Chandra}, Vedant and {Bauer}, Evan B. and {Fuller}, Jim and {Strader}, Jay and {Chomiuk}, Laura and {Naidu}, Rohan P. and {Caiazzo}, Ilaria and {Rodriguez}, Antonio C. and {Nagarajan}, Pranav and {Yamaguchi}, Natsuko and {Vanderbosch}, Zachary P. and {Roulston}, Benjamin R. and {G{\"a}nsicke}, Boris and {Han}, Jiwon Jesse and {Burdge}, Kevin B. and {Filippenko}, Alexei V. and {Brink}, Thomas G. and {Zheng}, WeiKang},
        title = "{The fastest stars in the Galaxy}",
      journal = {The Open Journal of Astrophysics},
         year = 2023,
        month = jul,
       volume = {6},
          eid = {28},
        pages = {28},
          doi = {10.21105/astro.2306.03914},
archivePrefix = {arXiv},
       eprint = {2306.03914},
 primaryClass = {astro-ph.SR},
       adsurl = {https://ui.adsabs.harvard.edu/abs/2023OJAp....6E..28E}
}

@ARTICLE{Raddi2018,
       author = {{Raddi}, R. and {Hollands}, M.~A. and {G{\"a}nsicke}, B.~T. and {Townsley}, D.~M. and {Hermes}, J.~J. and {Gentile Fusillo}, N.~P. and {Koester}, D.},
        title = "{Anatomy of the hyper-runaway star LP 40-365 with Gaia}",
      journal = {\mnras},
         year = 2018,
        month = sep,
       volume = {479},
       number = {1},
        pages = {L96-L101},
          doi = {10.1093/mnrasl/sly103},
archivePrefix = {arXiv},
       eprint = {1804.09677},
 primaryClass = {astro-ph.SR},
       adsurl = {https://ui.adsabs.harvard.edu/abs/2018MNRAS.479L..96R}
}

@ARTICLE{Vennes2017,
       author = {{Vennes}, S. and {Nemeth}, P. and {Kawka}, A. and {Thorstensen}, J.~R. and {Khalack}, V. and {Ferrario}, L. and {Alper}, E.~H.},
        title = "{An unusual white dwarf star may be a surviving remnant of a subluminous Type Ia supernova}",
      journal = {Science},
         year = 2017,
        month = aug,
       volume = {357},
       number = {6352},
        pages = {680-683},
          doi = {10.1126/science.aam8378},
archivePrefix = {arXiv},
       eprint = {1708.05568},
 primaryClass = {astro-ph.SR},
       adsurl = {https://ui.adsabs.harvard.edu/abs/2017Sci...357..680V}
}

@ARTICLE{Shen2025,
       author = {{Shen}, Ken J.},
        title = "{The Evolution of Hypervelocity Supernova Survivors and the Outcomes of Interacting Double White Dwarf Binaries}",
      journal = {\apj},
         year = 2025,
        month = mar,
       volume = {982},
       number = {1},
          eid = {6},
        pages = {6},
          doi = {10.3847/1538-4357/adb42e},
archivePrefix = {arXiv},
       eprint = {2502.04451},
 primaryClass = {astro-ph.SR},
       adsurl = {https://ui.adsabs.harvard.edu/abs/2025ApJ...982....6S}
}

@ARTICLE{Shen2024,
       author = {{Shen}, Ken J. and {Boos}, Samuel J. and {Townsley}, Dean M.},
        title = "{Almost All Carbon/Oxygen White Dwarfs Can Host Double Detonations}",
      journal = {\apj},
         year = 2024,
        month = nov,
       volume = {975},
       number = {1},
          eid = {127},
        pages = {127},
          doi = {10.3847/1538-4357/ad7379},
archivePrefix = {arXiv},
       eprint = {2405.19417},
 primaryClass = {astro-ph.SR},
       adsurl = {https://ui.adsabs.harvard.edu/abs/2024ApJ...975..127S}
}

@ARTICLE{Li2023,
       author = {{Li}, Zhenwei and {Chen}, Xuefei and {Ge}, Hongwei and {Chen}, Hai-Liang and {Han}, Zhanwen},
        title = "{Influence of a mass transfer stability criterion on double white dwarf populations}",
      journal = {\aap},
         year = 2023,
        month = jan,
       volume = {669},
          eid = {A82},
        pages = {A82},
          doi = {10.1051/0004-6361/202243893},
archivePrefix = {arXiv},
       eprint = {2211.01861},
 primaryClass = {astro-ph.SR},
       adsurl = {https://ui.adsabs.harvard.edu/abs/2023A&A...669A..82L}
}

@ARTICLE{vanZeist2025,
       author = {{van Zeist}, Wouter G.~J. and {van Roestel}, Jan and {Nelemans}, Gijs and {Eldridge}, Jan J. and {Korol}, Valeriya and {Toonen}, Silvia},
        title = "{Comparing population synthesis models of compact double white dwarfs to electromagnetic observations}",
      journal = {\aap},
         year = 2025,
        month = jul,
       volume = {699},
          eid = {A172},
        pages = {A172},
          doi = {10.1051/0004-6361/202554302},
archivePrefix = {arXiv},
       eprint = {2505.20953},
 primaryClass = {astro-ph.SR},
       adsurl = {https://ui.adsabs.harvard.edu/abs/2025A&A...699A.172V}
}

@ARTICLE{Tang2024,
       author = {{Tang}, P. and {Eldridge}, J.~J. and {Meyer}, R. and {Lamberts}, A. and {Boileau}, G. and {van Zeist}, W.~G.~J.},
        title = "{Predicting gravitational wave signals from BPASS white dwarf binary and black hole binary populations of a Milky Way-like galaxy model for LISA}",
      journal = {\mnras},
         year = 2024,
        month = nov,
       volume = {534},
       number = {3},
        pages = {1707-1728},
          doi = {10.1093/mnras/stae2154},
archivePrefix = {arXiv},
       eprint = {2405.20484},
 primaryClass = {astro-ph.GA},
       adsurl = {https://ui.adsabs.harvard.edu/abs/2024MNRAS.534.1707T}
}

@ARTICLE{Delfavero2025,
       author = {{Delfavero}, Vera and {Breivik}, Katelyn and {Thiele}, Sarah and {O'Shaughnessy}, Richard and {Baker}, John G.},
        title = "{Recovering Injected Astrophysics from the LISA Double White Dwarf Binaries}",
      journal = {\apj},
         year = 2025,
        month = mar,
       volume = {981},
       number = {1},
          eid = {66},
        pages = {66},
          doi = {10.3847/1538-4357/ada9e2},
archivePrefix = {arXiv},
       eprint = {2409.15230},
 primaryClass = {gr-qc},
       adsurl = {https://ui.adsabs.harvard.edu/abs/2025ApJ...981...66D}
}

@ARTICLE{DiCarlo2024,
       author = {{Di Carlo}, Ugo Niccol{\`o} and {Agrawal}, Poojan and {Rodriguez}, Carl L. and {Breivik}, Katelyn},
        title = "{Young Star Clusters Dominate the Production of Detached Black Hole─Star Binaries}",
      journal = {\apj},
         year = 2024,
        month = apr,
       volume = {965},
       number = {1},
          eid = {22},
        pages = {22},
          doi = {10.3847/1538-4357/ad2f2c},
archivePrefix = {arXiv},
       eprint = {2306.13121},
 primaryClass = {astro-ph.GA},
       adsurl = {https://ui.adsabs.harvard.edu/abs/2024ApJ...965...22D}
}

@ARTICLE{Chawla2022,
       author = {{Chawla}, Chirag and {Chatterjee}, Sourav and {Breivik}, Katelyn and {Moorthy}, Chaithanya Krishna and {Andrews}, Jeff J. and {Sanderson}, Robyn E.},
        title = "{Gaia May Detect Hundreds of Well-characterized Stellar Black Holes}",
      journal = {\apj},
         year = 2022,
        month = jun,
       volume = {931},
       number = {2},
          eid = {107},
        pages = {107},
          doi = {10.3847/1538-4357/ac60a5},
archivePrefix = {arXiv},
       eprint = {2110.05979},
 primaryClass = {astro-ph.GA},
       adsurl = {https://ui.adsabs.harvard.edu/abs/2022ApJ...931..107C}
}

@ARTICLE{Lamberts2018,
       author = {{Lamberts}, A. and {Garrison-Kimmel}, S. and {Hopkins}, P.~F. and {Quataert}, E. and {Bullock}, J.~S. and {Faucher-Gigu{\`e}re}, C.-A. and {Wetzel}, A. and {Kere{\v{s}}}, D. and {Drango}, K. and {Sanderson}, R.~E.},
        title = "{Predicting the binary black hole population of the Milky Way with cosmological simulations}",
      journal = {\mnras},
         year = 2018,
        month = oct,
       volume = {480},
       number = {2},
        pages = {2704-2718},
          doi = {10.1093/mnras/sty2035},
archivePrefix = {arXiv},
       eprint = {1801.03099},
 primaryClass = {astro-ph.GA},
       adsurl = {https://ui.adsabs.harvard.edu/abs/2018MNRAS.480.2704L}
}

@ARTICLE{Rajamuthukumar2025,
       author = {{Rajamuthukumar}, Abinaya Swaruba and {Korol}, Valeriya and {Stegmann}, Jakob and {Preece}, Holly and {Pakmor}, R{\"u}diger and {Justham}, Stephen and {Toonen}, Silvia and {de Mink}, Selma E.},
        title = "{The role of triple evolution in the formation of LISA double white dwarfs}",
      journal = {\aap},
         year = 2025,
        month = dec,
       volume = {704},
          eid = {A156},
        pages = {A156},
          doi = {10.1051/0004-6361/202554277},
archivePrefix = {arXiv},
       eprint = {2502.09607},
 primaryClass = {astro-ph.SR},
       adsurl = {https://ui.adsabs.harvard.edu/abs/2025A&A...704A.156R}
}

@ARTICLE{Rodriguez-Segovia2025,
       author = {{Rodr{\'\i}guez-Segovia}, Nicol{\'a}s and {Ruiter}, Ashley J.},
        title = "{Population synthesis of hot subdwarf B stars with COMPAS: on the observed Galactic population}",
      journal = {\mnras},
         year = 2025,
        month = jun,
       volume = {539},
       number = {4},
        pages = {3273-3284},
          doi = {10.1093/mnras/staf710},
archivePrefix = {arXiv},
       eprint = {2505.05791},
 primaryClass = {astro-ph.SR},
       adsurl = {https://ui.adsabs.harvard.edu/abs/2025MNRAS.539.3273R}
}

@ARTICLE{Dawson2024,
       author = {{Dawson}, H. and {Geier}, S. and {Heber}, U. and {Pelisoli}, I. and {Dorsch}, M. and {Schaffenroth}, V. and {Reindl}, N. and {Culpan}, R. and {Pritzkuleit}, M. and {Vos}, J. and {Soemitro}, A.~A. and {Roth}, M.~M. and {Schneider}, D. and {Uzundag}, M. and {Vu{\v{c}}kovi{\'c}}, M. and {Antunes Amaral}, L. and {Istrate}, A.~G. and {Justham}, S. and {{\O}stensen}, R.~H. and {Telting}, J.~H. and {Djupvik}, A.~A. and {Raddi}, R. and {Green}, E.~M. and {Jeffery}, C.~S. and {Kepler}, S.~O. and {Munday}, J. and {Steinmetz}, T. and {Kupfer}, T.},
        title = "{A 500 pc volume-limited sample of hot subluminous stars. I. Space density, scale height, and population properties}",
      journal = {\aap},
         year = 2024,
        month = jun,
       volume = {686},
          eid = {A25},
        pages = {A25},
          doi = {10.1051/0004-6361/202348319},
archivePrefix = {arXiv},
       eprint = {2403.15513},
 primaryClass = {astro-ph.SR},
       adsurl = {https://ui.adsabs.harvard.edu/abs/2024A&A...686A..25D}
}

@ARTICLE{Das2025,
       author = {{Das}, Priyam and {Seitenzahl}, Ivo R. and {Ruiter}, Ashley J. and {R{\"o}pke}, Friedrich K. and {Pakmor}, R{\"u}diger and {Vogt}, Fr{\'e}d{\'e}ric P.~A. and {Collins}, Christine E. and {Ghavamian}, Parviz and {Sim}, Stuart A. and {Williams}, Brian J. and {Taubenberger}, Stefan and {Laming}, J. Martin and {Suherli}, Janette and {Sutherland}, Ralph and {Rodr{\'\i}guez-Segovia}, Nicol{\'a}s},
        title = "{Calcium in a supernova remnant as a fingerprint of a sub-Chandrasekhar-mass explosion}",
      journal = {Nature Astronomy},
         year = 2025,
        month = sep,
       volume = {9},
        pages = {1356-1365},
          doi = {10.1038/s41550-025-02589-5},
       adsurl = {https://ui.adsabs.harvard.edu/abs/2025NatAs...9.1356D}
}

@ARTICLE{Han2002,
       author = {{Han}, Z. and {Podsiadlowski}, Ph. and {Maxted}, P.~F.~L. and {Marsh}, T.~R. and {Ivanova}, N.},
        title = "{The origin of subdwarf B stars - I. The formation channels}",
      journal = {\mnras},
         year = 2002,
        month = oct,
       volume = {336},
       number = {2},
        pages = {449-466},
          doi = {10.1046/j.1365-8711.2002.05752.x},
archivePrefix = {arXiv},
       eprint = {astro-ph/0206130},
 primaryClass = {astro-ph},
       adsurl = {https://ui.adsabs.harvard.edu/abs/2002MNRAS.336..449H}
}

@ARTICLE{DCruz1996,
       author = {{D'Cruz}, Noella L. and {Dorman}, Ben and {Rood}, Robert T. and {O'Connell}, Robert W.},
        title = "{The Origin of Extreme Horizontal Branch Stars}",
      journal = {\apj},
         year = 1996,
        month = jul,
       volume = {466},
        pages = {359},
          doi = {10.1086/177515},
archivePrefix = {arXiv},
       eprint = {astro-ph/9511017},
 primaryClass = {astro-ph},
       adsurl = {https://ui.adsabs.harvard.edu/abs/1996ApJ...466..359D}
}

@ARTICLE{OstrikerShetty2011,
       author = {{Ostriker}, Eve C. and {Shetty}, Rahul},
        title = "{Maximally Star-forming Galactic Disks. I. Starburst Regulation Via Feedback-driven Turbulence}",
      journal = {\apj},
         year = 2011,
        month = apr,
       volume = {731},
       number = {1},
          eid = {41},
        pages = {41},
          doi = {10.1088/0004-637X/731/1/41},
archivePrefix = {arXiv},
       eprint = {1102.1446},
 primaryClass = {astro-ph.CO},
       adsurl = {https://ui.adsabs.harvard.edu/abs/2011ApJ...731...41O}
}

@ARTICLE{Kobayashi2020,
       author = {{Kobayashi}, Chiaki and {Karakas}, Amanda I. and {Lugaro}, Maria},
        title = "{The Origin of Elements from Carbon to Uranium}",
      journal = {\apj},
         year = 2020,
        month = sep,
       volume = {900},
       number = {2},
          eid = {179},
        pages = {179},
          doi = {10.3847/1538-4357/abae65},
archivePrefix = {arXiv},
       eprint = {2008.04660},
 primaryClass = {astro-ph.GA},
       adsurl = {https://ui.adsabs.harvard.edu/abs/2020ApJ...900..179K}
}

@ARTICLE{Hopkins2014,
       author = {{Hopkins}, Philip F. and {Kere{\v{s}}}, Du{\v{s}}an and {O{\~n}orbe}, Jos{\'e} and {Faucher-Gigu{\`e}re}, Claude-Andr{\'e} and {Quataert}, Eliot and {Murray}, Norman and {Bullock}, James S.},
        title = "{Galaxies on FIRE (Feedback In Realistic Environments): stellar feedback explains cosmologically inefficient star formation}",
      journal = {\mnras},
         year = 2014,
        month = nov,
       volume = {445},
       number = {1},
        pages = {581-603},
          doi = {10.1093/mnras/stu1738},
archivePrefix = {arXiv},
       eprint = {1311.2073},
 primaryClass = {astro-ph.CO},
       adsurl = {https://ui.adsabs.harvard.edu/abs/2014MNRAS.445..581H}
}

@ARTICLE{SomervilleDave2015,
       author = {{Somerville}, Rachel S. and {Dav{\'e}}, Romeel},
        title = "{Physical Models of Galaxy Formation in a Cosmological Framework}",
      journal = {\araa},
         year = 2015,
        month = aug,
       volume = {53},
        pages = {51-113},
          doi = {10.1146/annurev-astro-082812-140951},
archivePrefix = {arXiv},
       eprint = {1412.2712},
 primaryClass = {astro-ph.GA},
       adsurl = {https://ui.adsabs.harvard.edu/abs/2015ARA&A..53...51S}
}

@ARTICLE{NaabOstriker2017,
       author = {{Naab}, Thorsten and {Ostriker}, Jeremiah P.},
        title = "{Theoretical Challenges in Galaxy Formation}",
      journal = {\araa},
         year = 2017,
        month = aug,
       volume = {55},
       number = {1},
        pages = {59-109},
          doi = {10.1146/annurev-astro-081913-040019},
archivePrefix = {arXiv},
       eprint = {1612.06891},
 primaryClass = {astro-ph.GA},
       adsurl = {https://ui.adsabs.harvard.edu/abs/2017ARA&A..55...59N}
}

@ARTICLE{Hopkins2018,
       author = {{Hopkins}, Philip F. and {Wetzel}, Andrew and {Kere{\v{s}}}, Du{\v{s}}an and {Faucher-Gigu{\`e}re}, Claude-Andr{\'e} and {Quataert}, Eliot and {Boylan-Kolchin}, Michael and {Murray}, Norman and {Hayward}, Christopher C. and {Garrison-Kimmel}, Shea and {Hummels}, Cameron and {Feldmann}, Robert and {Torrey}, Paul and {Ma}, Xiangcheng and {Angl{\'e}s-Alc{\'a}zar}, Daniel and {Su}, Kung-Yi and {Orr}, Matthew and {Schmitz}, Denise and {Escala}, Ivanna and {Sanderson}, Robyn and {Grudi{\'c}}, Michael Y. and {Hafen}, Zachary and {Kim}, Ji-Hoon and {Fitts}, Alex and {Bullock}, James S. and {Wheeler}, Coral and {Chan}, T.~K. and {Elbert}, Oliver D. and {Narayanan}, Desika},
        title = "{FIRE-2 simulations: physics versus numerics in galaxy formation}",
      journal = {\mnras},
         year = 2018,
        month = oct,
       volume = {480},
       number = {1},
        pages = {800-863},
          doi = {10.1093/mnras/sty1690},
archivePrefix = {arXiv},
       eprint = {1702.06148},
 primaryClass = {astro-ph.GA},
       adsurl = {https://ui.adsabs.harvard.edu/abs/2018MNRAS.480..800H}
}

@ARTICLE{Schwab2019,
       author = {{Schwab}, Josiah},
        title = "{Evolutionary Models for R Coronae Borealis Stars}",
      journal = {\apj},
         year = 2019,
        month = nov,
       volume = {885},
       number = {1},
          eid = {27},
        pages = {27},
          doi = {10.3847/1538-4357/ab425d},
archivePrefix = {arXiv},
       eprint = {1909.02569},
 primaryClass = {astro-ph.SR},
       adsurl = {https://ui.adsabs.harvard.edu/abs/2019ApJ...885...27S}
}

@ARTICLE{Heber2016,
       author = {{Heber}, U.},
        title = "{Hot Subluminous Stars}",
      journal = {\pasp},
         year = 2016,
        month = aug,
       volume = {128},
       number = {966},
        pages = {082001},
          doi = {10.1088/1538-3873/128/966/082001},
archivePrefix = {arXiv},
       eprint = {1604.07749},
 primaryClass = {astro-ph.SR},
       adsurl = {https://ui.adsabs.harvard.edu/abs/2016PASP..128h2001H}
}

@ARTICLE{SharaHurley2002,
       author = {{Shara}, Michael M. and {Hurley}, Jarrod R.},
        title = "{Star Clusters as Type Ia Supernova Factories}",
      journal = {\apj},
         year = 2002,
        month = jun,
       volume = {571},
       number = {2},
        pages = {830-842},
          doi = {10.1086/340062},
archivePrefix = {arXiv},
       eprint = {astro-ph/0202179},
 primaryClass = {astro-ph},
       adsurl = {https://ui.adsabs.harvard.edu/abs/2002ApJ...571..830S}
}

@ARTICLE{Claeys2014,
       author = {{Claeys}, J.~S.~W. and {Pols}, O.~R. and {Izzard}, R.~G. and {Vink}, J. and {Verbunt}, F.~W.~M.},
        title = "{Theoretical uncertainties of the Type Ia supernova rate}",
      journal = {\aap},
         year = 2014,
        month = mar,
       volume = {563},
          eid = {A83},
        pages = {A83},
          doi = {10.1051/0004-6361/201322714},
archivePrefix = {arXiv},
       eprint = {1401.2895},
 primaryClass = {astro-ph.SR},
       adsurl = {https://ui.adsabs.harvard.edu/abs/2014A&A...563A..83C}
}

@ARTICLE{Camacho2014,
       author = {{Camacho}, Judit and {Torres}, Santiago and {Garc{\'\i}a-Berro}, Enrique and {Zorotovic}, M{\'o}nica and {Schreiber}, Matthias R. and {Rebassa-Mansergas}, Alberto and {Nebot G{\'o}mez-Mor{\'a}n}, Ada and {G{\"a}nsicke}, Boris T.},
        title = "{Monte Carlo simulations of post-common-envelope white dwarf + main sequence binaries: comparison with the SDSS DR7 observed sample}",
      journal = {\aap},
         year = 2014,
        month = jun,
       volume = {566},
          eid = {A86},
        pages = {A86},
          doi = {10.1051/0004-6361/201323052},
archivePrefix = {arXiv},
       eprint = {1404.5464},
 primaryClass = {astro-ph.GA},
       adsurl = {https://ui.adsabs.harvard.edu/abs/2014A&A...566A..86C}
}

@ARTICLE{VerbuntRappaport1988,
       author = {{Verbunt}, Frank and {Rappaport}, Saul},
        title = "{Mass Transfer Instabilities Due to Angular Momentum Flows in Close Binaries}",
      journal = {\apj},
         year = 1988,
        month = sep,
       volume = {332},
        pages = {193},
          doi = {10.1086/166645},
       adsurl = {https://ui.adsabs.harvard.edu/abs/1988ApJ...332..193V}
}

@ARTICLE{Hansen2013,
       author = {{Hansen}, B.~M.~S. and {Kalirai}, J.~S. and {Anderson}, J. and {Dotter}, A. and {Richer}, H.~B. and {Rich}, R.~M. and {Shara}, M.~M. and {Fahlman}, G.~G. and {Hurley}, J.~R. and {King}, I.~R. and {Reitzel}, D. and {Stetson}, P.~B.},
        title = "{An age difference of two billion years between a metal-rich and a metal-poor globular cluster}",
      journal = {\nat},
         year = 2013,
        month = aug,
       volume = {500},
       number = {7460},
        pages = {51-53},
          doi = {10.1038/nature12334},
archivePrefix = {arXiv},
       eprint = {1308.0032},
 primaryClass = {astro-ph.SR},
       adsurl = {https://ui.adsabs.harvard.edu/abs/2013Natur.500...51H}
}

@ARTICLE{Cool1996,
       author = {{Cool}, Adrienne M. and {Piotto}, Giampaolo and {King}, Ivan R.},
        title = "{The Main Sequence and a White Dwarf Sequence in the Globular Cluster NGC 6397}",
      journal = {\apj},
         year = 1996,
        month = sep,
       volume = {468},
        pages = {655},
          doi = {10.1086/177723},
       adsurl = {https://ui.adsabs.harvard.edu/abs/1996ApJ...468..655C}
}

@ARTICLE{Renzini1996,
       author = {{Renzini}, Alvio and {Bragaglia}, Angela and {Ferraro}, Francesco R. and {Gilmozzi}, Roberto and {Ortolani}, Sergio and {Holberg}, J.~B. and {Liebert}, James and {Wesemael}, F. and {Bohlin}, Ralph C.},
        title = "{The White Dwarf Distance to the Globular Cluster NGC 6752 (and Its Age) with the Hubble Space Telescope}",
      journal = {\apjl},
         year = 1996,
        month = jul,
       volume = {465},
        pages = {L23},
          doi = {10.1086/310128},
       adsurl = {https://ui.adsabs.harvard.edu/abs/1996ApJ...465L..23R}
}

@ARTICLE{Richer1995,
       author = {{Richer}, Harvey B. and {Fahlman}, Gregory G. and {Ibata}, Rodrigo A. and {Stetson}, Peter B. and {Bell}, Roger A. and {Bolte}, Michael and {Bond}, Howard E. and {Harris}, William E. and {Hesser}, James E. and {Mandushev}, Georgi and {Pryor}, Carlton and {Vandenberg}, Don A.},
        title = "{Hubble Space Telescope Observations of White Dwarfs in the Globular Cluster M4}",
      journal = {\apjl},
         year = 1995,
        month = sep,
       volume = {451},
        pages = {L17},
          doi = {10.1086/309674},
archivePrefix = {arXiv},
       eprint = {astro-ph/9507109},
 primaryClass = {astro-ph},
       adsurl = {https://ui.adsabs.harvard.edu/abs/1995ApJ...451L..17R}
}

@ARTICLE{Rao2025,
       author = {{Rao}, Aryamann and {Ye}, Claire S. and {Fishbach}, Maya},
        title = "{Predicting the Rate of Fast Radio Bursts in Globular Clusters from Binary Black Hole Observations}",
      journal = {\apjl},
         year = 2025,
        month = jan,
       volume = {979},
       number = {1},
          eid = {L12},
        pages = {L12},
          doi = {10.3847/2041-8213/ad9f2e},
archivePrefix = {arXiv},
       eprint = {2409.20564},
 primaryClass = {astro-ph.HE},
       adsurl = {https://ui.adsabs.harvard.edu/abs/2025ApJ...979L..12R}
}

@ARTICLE{Metzger2021,
       author = {{Metzger}, Brian D. and {Zenati}, Yossef and {Chomiuk}, Laura and {Shen}, Ken J. and {Strader}, Jay},
        title = "{Transients from the Cataclysmic Deaths of Cataclysmic Variables}",
      journal = {\apj},
         year = 2021,
        month = dec,
       volume = {923},
       number = {1},
          eid = {100},
        pages = {100},
          doi = {10.3847/1538-4357/ac2a39},
archivePrefix = {arXiv},
       eprint = {2108.04305},
 primaryClass = {astro-ph.SR},
       adsurl = {https://ui.adsabs.harvard.edu/abs/2021ApJ...923..100M}
}

@ARTICLE{Geller2019,
       author = {{Geller}, Aaron M. and {Leigh}, Nathan W.~C. and {Giersz}, Mirek and {Kremer}, Kyle and {Rasio}, Frederic A.},
        title = "{In Search of the Thermal Eccentricity Distribution}",
      journal = {\apj},
         year = 2019,
        month = feb,
       volume = {872},
       number = {2},
          eid = {165},
        pages = {165},
          doi = {10.3847/1538-4357/ab0214},
archivePrefix = {arXiv},
       eprint = {1902.00019},
 primaryClass = {astro-ph.SR},
       adsurl = {https://ui.adsabs.harvard.edu/abs/2019ApJ...872..165G}
}

@ARTICLE{ToonenNelemans2013,
       author = {{Toonen}, S. and {Nelemans}, G.},
        title = "{The effect of common-envelope evolution on the visible population of post-common-envelope binaries}",
      journal = {\aap},
         year = 2013,
        month = sep,
       volume = {557},
          eid = {A87},
        pages = {A87},
          doi = {10.1051/0004-6361/201321753},
archivePrefix = {arXiv},
       eprint = {1309.0327},
 primaryClass = {astro-ph.SR},
       adsurl = {https://ui.adsabs.harvard.edu/abs/2013A&A...557A..87T}
}

@ARTICLE{Yamaguchi2024,
       author = {{Yamaguchi}, Natsuko and {El-Badry}, Kareem and {Fuller}, Jim and {Latham}, David W. and {Cargile}, Phillip A. and {Mazeh}, Tsevi and {Shahaf}, Sahar and {Bieryla}, Allyson and {Buchhave}, Lars A. and {Hobson}, Melissa},
        title = "{Wide post-common envelope binaries containing ultramassive white dwarfs: evidence for efficient envelope ejection in massive asymptotic giant branch stars}",
      journal = {\mnras},
         year = 2024,
        month = feb,
       volume = {527},
       number = {4},
        pages = {11719-11739},
          doi = {10.1093/mnras/stad4005},
archivePrefix = {arXiv},
       eprint = {2309.15905},
 primaryClass = {astro-ph.SR},
       adsurl = {https://ui.adsabs.harvard.edu/abs/2024MNRAS.52711719Y}
}

@ARTICLE{Scherbak2023,
       author = {{Scherbak}, Peter and {Fuller}, Jim},
        title = "{White dwarf binaries suggest a common envelope efficiency {\ensuremath{\alpha}}   1/3}",
      journal = {\mnras},
         year = 2023,
        month = jan,
       volume = {518},
       number = {3},
        pages = {3966-3984},
          doi = {10.1093/mnras/stac3313},
archivePrefix = {arXiv},
       eprint = {2211.02036},
 primaryClass = {astro-ph.SR},
       adsurl = {https://ui.adsabs.harvard.edu/abs/2023MNRAS.518.3966S}
}

@ARTICLE{zorotovic2010,
       author = {{Zorotovic}, M. and {Schreiber}, M.~R. and {G{\"a}nsicke}, B.~T. and {Nebot G{\'o}mez-Mor{\'a}n}, A.},
        title = "{Post-common-envelope binaries from SDSS. IX: Constraining the common-envelope efficiency}",
      journal = {\aap},
         year = 2010,
        month = sep,
       volume = {520},
          eid = {A86},
        pages = {A86},
          doi = {10.1051/0004-6361/200913658},
archivePrefix = {arXiv},
       eprint = {1006.1621},
 primaryClass = {astro-ph.SR},
       adsurl = {https://ui.adsabs.harvard.edu/abs/2010A&A...520A..86Z}
}

@ARTICLE{demarco2009,
       author = {{De Marco}, Orsola},
        title = "{The Origin and Shaping of Planetary Nebulae: Putting the Binary Hypothesis to the Test}",
      journal = {\pasp},
         year = 2009,
        month = apr,
       volume = {121},
       number = {878},
        pages = {316},
          doi = {10.1086/597765},
archivePrefix = {arXiv},
       eprint = {0902.1137},
 primaryClass = {astro-ph.GA},
       adsurl = {https://ui.adsabs.harvard.edu/abs/2009PASP..121..316D}
}

@ARTICLE{Lorimer2008,
       author = {{Lorimer}, Duncan R.},
        title = "{Binary and Millisecond Pulsars}",
      journal = {Living Reviews in Relativity},
         year = 2008,
        month = dec,
       volume = {11},
       number = {1},
          eid = {8},
        pages = {8},
          doi = {10.12942/lrr-2008-8},
archivePrefix = {arXiv},
       eprint = {0811.0762},
 primaryClass = {astro-ph},
       adsurl = {https://ui.adsabs.harvard.edu/abs/2008LRR....11....8L}
}

@ARTICLE{Green2025,
       author = {{Green}, Matthew J. and {van Roestel}, Jan and {Wong}, Tin Long Sunny},
        title = "{A catalogue of ultracompact mass-transferring white dwarf binaries}",
      journal = {\aap},
         year = 2025,
        month = aug,
       volume = {700},
          eid = {A107},
        pages = {A107},
          doi = {10.1051/0004-6361/202554925},
archivePrefix = {arXiv},
       eprint = {2505.10535},
 primaryClass = {astro-ph.SR},
       adsurl = {https://ui.adsabs.harvard.edu/abs/2025A&A...700A.107G}
}

@ARTICLE{Solheim2010,
       author = {{Solheim}, J.-E.},
        title = "{AM CVn Stars: Status and Challenges}",
      journal = {\pasp},
         year = 2010,
        month = oct,
       volume = {122},
       number = {896},
        pages = {1133},
          doi = {10.1086/656680},
       adsurl = {https://ui.adsabs.harvard.edu/abs/2010PASP..122.1133S}
}

@ARTICLE{vanRoestel2022,
       author = {{van Roestel}, J. and {Kupfer}, T. and {Green}, M.~J. and {Wong}, T.~L.~S. and {Bildsten}, L. and {Burdge}, K. and {Prince}, T. and {Marsh}, T.~R. and {Szkody}, P. and {Fremling}, C. and {Graham}, M.~J. and {Dhillon}, V.~S. and {Littlefair}, S.~P. and {Bellm}, E.~C. and {Coughlin}, M. and {Duev}, D.~A. and {Goldstein}, D.~A. and {Laher}, R.~R. and {Rusholme}, B. and {Riddle}, R. and {Dekany}, R. and {Kulkarni}, S.~R.},
        title = "{Discovery and characterization of five new eclipsing AM CVn systems}",
      journal = {\mnras},
         year = 2022,
        month = jun,
       volume = {512},
       number = {4},
        pages = {5440-5461},
          doi = {10.1093/mnras/stab2421},
archivePrefix = {arXiv},
       eprint = {2107.07573},
 primaryClass = {astro-ph.SR},
       adsurl = {https://ui.adsabs.harvard.edu/abs/2022MNRAS.512.5440V}
}

@ARTICLE{Ramsay2018,
       author = {{Ramsay}, G. and {Green}, M.~J. and {Marsh}, T.~R. and {Kupfer}, T. and {Breedt}, E. and {Korol}, V. and {Groot}, P.~J. and {Knigge}, C. and {Nelemans}, G. and {Steeghs}, D. and {Woudt}, P. and {Aungwerojwit}, A.},
        title = "{Physical properties of AM CVn stars: New insights from Gaia DR2}",
      journal = {\aap},
         year = 2018,
        month = dec,
       volume = {620},
          eid = {A141},
        pages = {A141},
          doi = {10.1051/0004-6361/201834261},
archivePrefix = {arXiv},
       eprint = {1810.06548},
 primaryClass = {astro-ph.SR},
       adsurl = {https://ui.adsabs.harvard.edu/abs/2018A&A...620A.141R}
}

@ARTICLE{Toonen2018,
       author = {{Toonen}, S. and {Perets}, H.~B. and {Hamers}, A.~S.},
        title = "{Rate of WD-WD head-on collisions in isolated triples is too low to explain standard type Ia supernovae}",
      journal = {\aap},
         year = 2018,
        month = feb,
       volume = {610},
          eid = {A22},
        pages = {A22},
          doi = {10.1051/0004-6361/201731874},
archivePrefix = {arXiv},
       eprint = {1709.00422},
 primaryClass = {astro-ph.HE},
       adsurl = {https://ui.adsabs.harvard.edu/abs/2018A&A...610A..22T}
}

@ARTICLE{Naoz2016,
       author = {{Naoz}, Smadar},
        title = "{The Eccentric Kozai-Lidov Effect and Its Applications}",
      journal = {\araa},
         year = 2016,
        month = sep,
       volume = {54},
        pages = {441-489},
          doi = {10.1146/annurev-astro-081915-023315},
archivePrefix = {arXiv},
       eprint = {1601.07175},
 primaryClass = {astro-ph.EP},
       adsurl = {https://ui.adsabs.harvard.edu/abs/2016ARA&A..54..441N}
}

@ARTICLE{Lidov1962,
       author = {{Lidov}, M.~L.},
        title = "{The evolution of orbits of artificial satellites of planets under the action of gravitational perturbations of external bodies}",
      journal = {\planss},
         year = 1962,
        month = oct,
       volume = {9},
       number = {10},
        pages = {719-759},
          doi = {10.1016/0032-0633(62)90129-0},
       adsurl = {https://ui.adsabs.harvard.edu/abs/1962P&SS....9..719L}
}

@ARTICLE{Kozai1962,
       author = {{Kozai}, Yoshihide},
        title = "{Secular perturbations of asteroids with high inclination and eccentricity}",
      journal = {\aj},
         year = 1962,
        month = nov,
       volume = {67},
        pages = {591-598},
          doi = {10.1086/108790},
       adsurl = {https://ui.adsabs.harvard.edu/abs/1962AJ.....67..591K}
}

@ARTICLE{Shariat2025,
       author = {{Shariat}, Cheyanne and {El-Badry}, Kareem and {Naoz}, Smadar},
        title = "{10,000 Resolved Triples from Gaia: Empirical Constraints on Triple Star Populations}",
      journal = {\pasp},
         year = 2025,
        month = sep,
       volume = {137},
       number = {9},
          eid = {094201},
        pages = {094201},
          doi = {10.1088/1538-3873/adfb30},
archivePrefix = {arXiv},
       eprint = {2506.16513},
 primaryClass = {astro-ph.SR},
       adsurl = {https://ui.adsabs.harvard.edu/abs/2025PASP..137i4201S}
}

@INPROCEEDINGS{Offner2023,
       author = {{Offner}, S.~S.~R. and {Moe}, M. and {Kratter}, K.~M. and {Sadavoy}, S.~I. and {Jensen}, E.~L.~N. and {Tobin}, J.~J.},
        title = "{The Origin and Evolution of Multiple Star Systems}",
    booktitle = {Protostars and Planets VII},
         year = 2023,
       editor = {{Inutsuka}, S. and {Aikawa}, Y. and {Muto}, T. and {Tomida}, K. and {Tamura}, M.},
       series = {Astronomical Society of the Pacific Conference Series},
       volume = {534},
        month = jul,
        pages = {275},
          doi = {10.48550/arXiv.2203.10066},
archivePrefix = {arXiv},
       eprint = {2203.10066},
 primaryClass = {astro-ph.SR},
       adsurl = {https://ui.adsabs.harvard.edu/abs/2023ASPC..534..275O}
}

@ARTICLE{MoeDiStefano2017,
       author = {{Moe}, Maxwell and {Di Stefano}, Rosanne},
        title = "{Mind Your Ps and Qs: The Interrelation between Period (P) and Mass-ratio (Q) Distributions of Binary Stars}",
      journal = {\apjs},
         year = 2017,
        month = jun,
       volume = {230},
       number = {2},
          eid = {15},
        pages = {15},
          doi = {10.3847/1538-4365/aa6fb6},
archivePrefix = {arXiv},
       eprint = {1606.05347},
 primaryClass = {astro-ph.SR},
       adsurl = {https://ui.adsabs.harvard.edu/abs/2017ApJS..230...15M}
}

@ARTICLE{Tokovinin2014,
       author = {{Tokovinin}, Andrei},
        title = "{From Binaries to Multiples. II. Hierarchical Multiplicity of F and G Dwarfs}",
      journal = {\aj},
         year = 2014,
        month = apr,
       volume = {147},
       number = {4},
          eid = {87},
        pages = {87},
          doi = {10.1088/0004-6256/147/4/87},
archivePrefix = {arXiv},
       eprint = {1401.6827},
 primaryClass = {astro-ph.SR},
       adsurl = {https://ui.adsabs.harvard.edu/abs/2014AJ....147...87T}
}

@ARTICLE{Raghavan2010,
       author = {{Raghavan}, Deepak and {McAlister}, Harold A. and {Henry}, Todd J. and {Latham}, David W. and {Marcy}, Geoffrey W. and {Mason}, Brian D. and {Gies}, Douglas R. and {White}, Russel J. and {ten Brummelaar}, Theo A.},
        title = "{A Survey of Stellar Families: Multiplicity of Solar-type Stars}",
      journal = {\apjs},
         year = 2010,
        month = sep,
       volume = {190},
       number = {1},
        pages = {1-42},
          doi = {10.1088/0067-0049/190/1/1},
archivePrefix = {arXiv},
       eprint = {1007.0414},
 primaryClass = {astro-ph.SR},
       adsurl = {https://ui.adsabs.harvard.edu/abs/2010ApJS..190....1R}
}

@ARTICLE{Fang2018,
       author = {{Fang}, Xiao and {Thompson}, Todd A. and {Hirata}, Christopher M.},
        title = "{Dynamics of quadruple systems composed of two binaries: stars, white dwarfs, and implications for Ia supernovae}",
      journal = {\mnras},
         year = 2018,
        month = may,
       volume = {476},
       number = {3},
        pages = {4234-4262},
          doi = {10.1093/mnras/sty472},
archivePrefix = {arXiv},
       eprint = {1709.08682},
 primaryClass = {astro-ph.HE},
       adsurl = {https://ui.adsabs.harvard.edu/abs/2018MNRAS.476.4234F}
}

@ARTICLE{AntogniniThompson2014,
       author = {{Antognini}, Joe M. and {Shappee}, Benjamin J. and {Thompson}, Todd A. and {Amaro-Seoane}, Pau},
        title = "{Rapid eccentricity oscillations and the mergers of compact objects in hierarchical triples}",
      journal = {\mnras},
         year = 2014,
        month = mar,
       volume = {439},
       number = {1},
        pages = {1079-1091},
          doi = {10.1093/mnras/stu039},
archivePrefix = {arXiv},
       eprint = {1308.5682},
 primaryClass = {astro-ph.HE},
       adsurl = {https://ui.adsabs.harvard.edu/abs/2014MNRAS.439.1079A}
}

@ARTICLE{Thompson2011,
       author = {{Thompson}, Todd A.},
        title = "{Accelerating Compact Object Mergers in Triple Systems with the Kozai Resonance: A Mechanism for ``Prompt'' Type Ia Supernovae, Gamma-Ray Bursts, and Other Exotica}",
      journal = {\apj},
         year = 2011,
        month = nov,
       volume = {741},
       number = {2},
          eid = {82},
        pages = {82},
          doi = {10.1088/0004-637X/741/2/82},
archivePrefix = {arXiv},
       eprint = {1011.4322},
 primaryClass = {astro-ph.HE},
       adsurl = {https://ui.adsabs.harvard.edu/abs/2011ApJ...741...82T}
}

@ARTICLE{Sheriat2026,
       author = {{Shariat}, Cheyanne and {Ye}, Claire S. and {Naoz}, Smadar and {Rose}, Sanaea C.},
        title = "{Fast Radio Bursts from White Dwarf Binary Mergers: Isolated and Triple-induced Channels}",
      journal = {\apjl},
         year = 2026,
        month = mar,
       volume = {1000},
       number = {1},
          eid = {L17},
        pages = {L17},
          doi = {10.3847/2041-8213/ae4d17},
archivePrefix = {arXiv},
       eprint = {2511.18678},
 primaryClass = {astro-ph.HE},
       adsurl = {https://ui.adsabs.harvard.edu/abs/2026ApJ..1000L..17S}
}

@ARTICLE{WashabaughBregman2013,
       author = {{Washabaugh}, Pearce C. and {Bregman}, Joel N.},
        title = "{The Production Rate of SN Ia Events in Globular Clusters}",
      journal = {\apj},
         year = 2013,
        month = jan,
       volume = {762},
       number = {1},
          eid = {1},
        pages = {1},
          doi = {10.1088/0004-637X/762/1/1},
archivePrefix = {arXiv},
       eprint = {1205.0588},
 primaryClass = {astro-ph.HE},
       adsurl = {https://ui.adsabs.harvard.edu/abs/2013ApJ...762....1W}
}

@ARTICLE{VossNelemans2012,
       author = {{Voss}, R. and {Nelemans}, G.},
        title = "{Type Ia supernovae in globular clusters: observational upper limits}",
      journal = {\aap},
         year = 2012,
        month = mar,
       volume = {539},
          eid = {A77},
        pages = {A77},
          doi = {10.1051/0004-6361/201118222},
archivePrefix = {arXiv},
       eprint = {1111.6593},
 primaryClass = {astro-ph.HE},
       adsurl = {https://ui.adsabs.harvard.edu/abs/2012A&A...539A..77V}
}

@ARTICLE{Bregman2024,
       author = {{Bregman}, Joel N. and {Gnedin}, Oleg Y. and {Seitzer}, Patrick O. and {Qu}, Zhijie},
        title = "{A Type Ia Supernova near a Globular Cluster in the Early-type Galaxy NGC 5353}",
      journal = {\apjl},
         year = 2024,
        month = jun,
       volume = {968},
       number = {1},
          eid = {L6},
        pages = {L6},
          doi = {10.3847/2041-8213/ad498f},
archivePrefix = {arXiv},
       eprint = {2405.09701},
 primaryClass = {astro-ph.HE},
       adsurl = {https://ui.adsabs.harvard.edu/abs/2024ApJ...968L...6B}
}

@INPROCEEDINGS{Kremer2026,
       author = {{Kremer}, Kyle},
        title = "{Compact objects in globular clusters}",
    booktitle = {Encyclopedia of Astrophysics, Volume 3},
         year = 2026,
       volume = {3},
        month = jan,
        pages = {458-472},
          doi = {10.1016/B978-0-443-21439-4.00103-6},
archivePrefix = {arXiv},
       eprint = {2508.14308},
 primaryClass = {astro-ph.HE},
       adsurl = {https://ui.adsabs.harvard.edu/abs/2026enap....3..458K}
}

@ARTICLE{Bahramian2013,
       author = {{Bahramian}, Arash and {Heinke}, Craig O. and {Sivakoff}, Gregory R. and {Gladstone}, Jeanette C.},
        title = "{Stellar Encounter Rate in Galactic Globular Clusters}",
      journal = {\apj},
         year = 2013,
        month = apr,
       volume = {766},
       number = {2},
          eid = {136},
        pages = {136},
          doi = {10.1088/0004-637X/766/2/136},
archivePrefix = {arXiv},
       eprint = {1302.2549},
 primaryClass = {astro-ph.HE},
       adsurl = {https://ui.adsabs.harvard.edu/abs/2013ApJ...766..136B}
}

@ARTICLE{Katz1975,
       author = {{Katz}, J.~I.},
        title = "{Two kinds of stellar collapse}",
      journal = {\nat},
         year = 1975,
        month = feb,
       volume = {253},
       number = {5494},
        pages = {698-699},
          doi = {10.1038/253698a0},
       adsurl = {https://ui.adsabs.harvard.edu/abs/1975Natur.253..698K}
}

@ARTICLE{Lu2022,
       author = {{Lu}, Wenbin and {Beniamini}, Paz and {Kumar}, Pawan},
        title = "{Implications of a rapidly varying FRB in a globular cluster of M81}",
      journal = {\mnras},
         year = 2022,
        month = feb,
       volume = {510},
       number = {2},
        pages = {1867-1879},
          doi = {10.1093/mnras/stab3500},
archivePrefix = {arXiv},
       eprint = {2107.04059},
 primaryClass = {astro-ph.HE},
       adsurl = {https://ui.adsabs.harvard.edu/abs/2022MNRAS.510.1867L}
}

@ARTICLE{Kristen2022,
       author = {{Kirsten}, F. and {Marcote}, B. and {Nimmo}, K. and {Hessels}, J.~W.~T. and {Bhardwaj}, M. and {Tendulkar}, S.~P. and {Keimpema}, A. and {Yang}, J. and {Snelders}, M.~P. and {Scholz}, P. and {Pearlman}, A.~B. and {Law}, C.~J. and {Peters}, W.~M. and {Giroletti}, M. and {Paragi}, Z. and {Bassa}, C. and {Hewitt}, D.~M. and {Bach}, U. and {Bezrukovs}, V. and {Burgay}, M. and {Buttaccio}, S.~T. and {Conway}, J.~E. and {Corongiu}, A. and {Feiler}, R. and {Forss{\'e}n}, O. and {Gawro{\'n}ski}, M.~P. and {Karuppusamy}, R. and {Kharinov}, M.~A. and {Lindqvist}, M. and {Maccaferri}, G. and {Melnikov}, A. and {Ould-Boukattine}, O.~S. and {Possenti}, A. and {Surcis}, G. and {Wang}, N. and {Yuan}, J. and {Aggarwal}, K. and {Anna-Thomas}, R. and {Bower}, G.~C. and {Blaauw}, R. and {Burke-Spolaor}, S. and {Cassanelli}, T. and {Clarke}, T.~E. and {Fonseca}, E. and {Gaensler}, B.~M. and {Gopinath}, A. and {Kaspi}, V.~M. and {Kassim}, N. and {Lazio}, T.~J.~W. and {Leung}, C. and {Li}, D.~Z. and {Lin}, H.~H. and {Masui}, K.~W. and {Mckinven}, R. and {Michilli}, D. and {Mikhailov}, A.~G. and {Ng}, C. and {Orbidans}, A. and {Pen}, U.~L. and {Petroff}, E. and {Rahman}, M. and {Ransom}, S.~M. and {Shin}, K. and {Smith}, K.~M. and {Stairs}, I.~H. and {Vlemmings}, W.},
        title = "{A repeating fast radio burst source in a globular cluster}",
      journal = {\nat},
         year = 2022,
        month = feb,
       volume = {602},
       number = {7898},
        pages = {585-589},
          doi = {10.1038/s41586-021-04354-w},
archivePrefix = {arXiv},
       eprint = {2105.11445},
 primaryClass = {astro-ph.HE},
       adsurl = {https://ui.adsabs.harvard.edu/abs/2022Natur.602..585K}
}

@ARTICLE{MaozMannucci2012,
       author = {{Maoz}, D. and {Mannucci}, F.},
        title = "{Type-Ia Supernova Rates and the Progenitor Problem: A Review}",
      journal = {\pasa},
         year = 2012,
        month = jan,
       volume = {29},
       number = {4},
        pages = {447-465},
          doi = {10.1071/AS11052},
archivePrefix = {arXiv},
       eprint = {1111.4492},
 primaryClass = {astro-ph.CO},
       adsurl = {https://ui.adsabs.harvard.edu/abs/2012PASA...29..447M}
}

@ARTICLE{Maoz2014,
       author = {{Maoz}, Dan and {Mannucci}, Filippo and {Nelemans}, Gijs},
        title = "{Observational Clues to the Progenitors of Type Ia Supernovae}",
      journal = {\araa},
         year = 2014,
        month = aug,
       volume = {52},
        pages = {107-170},
          doi = {10.1146/annurev-astro-082812-141031},
archivePrefix = {arXiv},
       eprint = {1312.0628},
 primaryClass = {astro-ph.CO},
       adsurl = {https://ui.adsabs.harvard.edu/abs/2014ARA&A..52..107M}
}

@ARTICLE{Li2011,
       author = {{Li}, Weidong and {Chornock}, Ryan and {Leaman}, Jesse and {Filippenko}, Alexei V. and {Poznanski}, Dovi and {Wang}, Xiaofeng and {Ganeshalingam}, Mohan and {Mannucci}, Filippo},
        title = "{Nearby supernova rates from the Lick Observatory Supernova Search - III. The rate-size relation, and the rates as a function of galaxy Hubble type and colour}",
      journal = {\mnras},
         year = 2011,
        month = apr,
       volume = {412},
       number = {3},
        pages = {1473-1507},
          doi = {10.1111/j.1365-2966.2011.18162.x},
archivePrefix = {arXiv},
       eprint = {1006.4613},
 primaryClass = {astro-ph.SR},
       adsurl = {https://ui.adsabs.harvard.edu/abs/2011MNRAS.412.1473L}
}

@ARTICLE{Liu2023,
       author = {{Liu}, Zheng-Wei and {R{\"o}pke}, Friedrich K. and {Han}, Zhanwen},
        title = "{Type Ia Supernova Explosions in Binary Systems: A Review}",
      journal = {Research in Astronomy and Astrophysics},
         year = 2023,
        month = aug,
       volume = {23},
       number = {8},
          eid = {082001},
        pages = {082001},
          doi = {10.1088/1674-4527/acd89e},
archivePrefix = {arXiv},
       eprint = {2305.13305},
 primaryClass = {astro-ph.HE},
       adsurl = {https://ui.adsabs.harvard.edu/abs/2023RAA....23h2001L}
}

@ARTICLE{Munday2025,
       author = {{He}, Wenlang and {Zhou}, Ping and {Gjergo}, Eda and {Fu}, Xiaoting},
        title = "{Alien Type Ia Supernovae from the Milky Way Merger History and One Possible Candidate: Kepler's Supernova}",
      journal = {\apj},
         year = 2025,
        month = jun,
       volume = {986},
       number = {2},
          eid = {123},
        pages = {123},
          doi = {10.3847/1538-4357/add32b},
archivePrefix = {arXiv},
       eprint = {2505.03085},
 primaryClass = {astro-ph.HE},
       adsurl = {https://ui.adsabs.harvard.edu/abs/2025ApJ...986..123H}
}

@ARTICLE{Tauris2015,
       author = {{Tauris}, Thomas M. and {Langer}, Norbert and {Podsiadlowski}, Philipp},
        title = "{Ultra-stripped supernovae: progenitors and fate}",
      journal = {\mnras},
         year = 2015,
        month = aug,
       volume = {451},
       number = {2},
        pages = {2123-2144},
          doi = {10.1093/mnras/stv990},
archivePrefix = {arXiv},
       eprint = {1505.00270},
 primaryClass = {astro-ph.SR},
       adsurl = {https://ui.adsabs.harvard.edu/abs/2015MNRAS.451.2123T}
}

@ARTICLE{Han2003,
       author = {{Han}, Z. and {Podsiadlowski}, Ph. and {Maxted}, P.~F.~L. and {Marsh}, T.~R.},
        title = "{The origin of subdwarf B stars - II}",
      journal = {\mnras},
         year = 2003,
        month = may,
       volume = {341},
       number = {2},
        pages = {669-691},
          doi = {10.1046/j.1365-8711.2003.06451.x},
archivePrefix = {arXiv},
       eprint = {astro-ph/0301380},
 primaryClass = {astro-ph},
       adsurl = {https://ui.adsabs.harvard.edu/abs/2003MNRAS.341..669H}
}

@ARTICLE{ShenMoore2014,
       author = {{Shen}, Ken J. and {Moore}, Kevin},
        title = "{The Initiation and Propagation of Helium Detonations in White Dwarf Envelopes}",
      journal = {\apj},
         year = 2014,
        month = dec,
       volume = {797},
       number = {1},
          eid = {46},
        pages = {46},
          doi = {10.1088/0004-637X/797/1/46},
archivePrefix = {arXiv},
       eprint = {1409.3568},
 primaryClass = {astro-ph.HE},
       adsurl = {https://ui.adsabs.harvard.edu/abs/2014ApJ...797...46S}
}

@ARTICLE{RudermanSutherland1975,
       author = {{Ruderman}, M.~A. and {Sutherland}, P.~G.},
        title = "{Theory of pulsars: polar gaps, sparks, and coherent microwave radiation.}",
      journal = {\apj},
         year = 1975,
        month = feb,
       volume = {196},
        pages = {51-72},
          doi = {10.1086/153393},
       adsurl = {https://ui.adsabs.harvard.edu/abs/1975ApJ...196...51R}
}

@BOOK{ShapiroTeukolsky1983,
       author = {{Shapiro}, Stuart L. and {Teukolsky}, Saul A.},
        title = "{Black holes, white dwarfs and neutron stars. The physics of compact objects}",
         year = 1983,
          doi = {10.1002/9783527617661},
       adsurl = {https://ui.adsabs.harvard.edu/abs/1983bhwd.book.....S}
}

@ARTICLE{Fink2010,
       author = {{Fink}, M. and {R{\"o}pke}, F.~K. and {Hillebrandt}, W. and {Seitenzahl}, I.~R. and {Sim}, S.~A. and {Kromer}, M.},
        title = "{Double-detonation sub-Chandrasekhar supernovae: can minimum helium shell masses detonate the core?}",
      journal = {\aap},
         year = 2010,
        month = may,
       volume = {514},
          eid = {A53},
        pages = {A53},
          doi = {10.1051/0004-6361/200913892},
archivePrefix = {arXiv},
       eprint = {1002.2173},
 primaryClass = {astro-ph.SR},
       adsurl = {https://ui.adsabs.harvard.edu/abs/2010A&A...514A..53F}
}

@ARTICLE{WoosleyWeaver1994,
       author = {{Woosley}, S.~E. and {Weaver}, Thomas A.},
        title = "{Sub--Chandrasekhar Mass Models for Type IA Supernovae}",
      journal = {\apj},
         year = 1994,
        month = mar,
       volume = {423},
        pages = {371},
          doi = {10.1086/173813},
       adsurl = {https://ui.adsabs.harvard.edu/abs/1994ApJ...423..371W}
}

@ARTICLE{Drout2014,
       author = {{Drout}, M.~R. and {Chornock}, R. and {Soderberg}, A.~M. and {Sanders}, N.~E. and {McKinnon}, R. and {Rest}, A. and {Foley}, R.~J. and {Milisavljevic}, D. and {Margutti}, R. and {Berger}, E. and {Calkins}, M. and {Fong}, W. and {Gezari}, S. and {Huber}, M.~E. and {Kankare}, E. and {Kirshner}, R.~P. and {Leibler}, C. and {Lunnan}, R. and {Mattila}, S. and {Marion}, G.~H. and {Narayan}, G. and {Riess}, A.~G. and {Roth}, K.~C. and {Scolnic}, D. and {Smartt}, S.~J. and {Tonry}, J.~L. and {Burgett}, W.~S. and {Chambers}, K.~C. and {Hodapp}, K.~W. and {Jedicke}, R. and {Kaiser}, N. and {Magnier}, E.~A. and {Metcalfe}, N. and {Morgan}, J.~S. and {Price}, P.~A. and {Waters}, C.},
        title = "{Rapidly Evolving and Luminous Transients from Pan-STARRS1}",
      journal = {\apj},
         year = 2014,
        month = oct,
       volume = {794},
       number = {1},
          eid = {23},
        pages = {23},
          doi = {10.1088/0004-637X/794/1/23},
archivePrefix = {arXiv},
       eprint = {1405.3668},
 primaryClass = {astro-ph.HE},
       adsurl = {https://ui.adsabs.harvard.edu/abs/2014ApJ...794...23D}
}

@ARTICLE{Dessart2006,
       author = {{Dessart}, L. and {Burrows}, A. and {Ott}, C.~D. and {Livne}, E. and {Yoon}, S.-C. and {Langer}, N.},
        title = "{Multidimensional Simulations of the Accretion-induced Collapse of White Dwarfs to Neutron Stars}",
      journal = {\apj},
         year = 2006,
        month = jun,
       volume = {644},
       number = {2},
        pages = {1063-1084},
          doi = {10.1086/503626},
archivePrefix = {arXiv},
       eprint = {astro-ph/0601603},
 primaryClass = {astro-ph},
       adsurl = {https://ui.adsabs.harvard.edu/abs/2006ApJ...644.1063D}
}

@ARTICLE{Kitaura2006,
       author = {{Kitaura}, F.~S. and {Janka}, H.-Th. and {Hillebrandt}, W.},
        title = "{Explosions of O-Ne-Mg cores, the Crab supernova, and subluminous type II-P supernovae}",
      journal = {\aap},
         year = 2006,
        month = apr,
       volume = {450},
       number = {1},
        pages = {345-350},
          doi = {10.1051/0004-6361:20054703},
archivePrefix = {arXiv},
       eprint = {astro-ph/0512065},
 primaryClass = {astro-ph},
       adsurl = {https://ui.adsabs.harvard.edu/abs/2006A&A...450..345K}
}

@ARTICLE{Cheng2019,
       author = {{Cheng}, Sihao and {Cummings}, Jeffrey D. and {M{\'e}nard}, Brice},
        title = "{A Cooling Anomaly of High-mass White Dwarfs}",
      journal = {\apj},
         year = 2019,
        month = dec,
       volume = {886},
       number = {2},
          eid = {100},
        pages = {100},
          doi = {10.3847/1538-4357/ab4989},
archivePrefix = {arXiv},
       eprint = {1905.12710},
 primaryClass = {astro-ph.SR},
       adsurl = {https://ui.adsabs.harvard.edu/abs/2019ApJ...886..100C}
}

@ARTICLE{KoesterKepler2019,
       author = {{Koester}, D. and {Kepler}, S.~O.},
        title = "{Carbon-rich (DQ) white dwarfs in the Sloan Digital Sky Survey}",
      journal = {\aap},
         year = 2019,
        month = aug,
       volume = {628},
          eid = {A102},
        pages = {A102},
          doi = {10.1051/0004-6361/201935946},
archivePrefix = {arXiv},
       eprint = {1905.11174},
 primaryClass = {astro-ph.SR},
       adsurl = {https://ui.adsabs.harvard.edu/abs/2019A&A...628A.102K}
}

@ARTICLE{Coutu2019,
       author = {{Coutu}, S. and {Dufour}, P. and {Bergeron}, P. and {Blouin}, S. and {Loranger}, E. and {Allard}, N.~F. and {Dunlap}, B.~H.},
        title = "{Analysis of Helium-rich White Dwarfs Polluted by Heavy Elements in the Gaia Era}",
      journal = {\apj},
         year = 2019,
        month = nov,
       volume = {885},
       number = {1},
          eid = {74},
        pages = {74},
          doi = {10.3847/1538-4357/ab46b9},
archivePrefix = {arXiv},
       eprint = {1907.05932},
 primaryClass = {astro-ph.SR},
       adsurl = {https://ui.adsabs.harvard.edu/abs/2019ApJ...885...74C}
}

@INPROCEEDINGS{DunlapClemens2015,
       author = {{Dunlap}, Bart H. and {Clemens}, J.~C.},
        title = "{Hot DQ White Dwarf Stars as Failed Type Ia Supernovae}",
    booktitle = {19th European Workshop on White Dwarfs},
         year = 2015,
       editor = {{Dufour}, P. and {Bergeron}, P. and {Fontaine}, G.},
       series = {Astronomical Society of the Pacific Conference Series},
       volume = {493},
        month = jun,
        pages = {547},
       adsurl = {https://ui.adsabs.harvard.edu/abs/2015ASPC..493..547D}
}

@ARTICLE{Gvaramadze2019,
       author = {{Gvaramadze}, Vasilii V. and {Gr{\"a}fener}, G{\"o}tz and {Langer}, Norbert and {Maryeva}, Olga V. and {Kniazev}, Alexei Y. and {Moskvitin}, Alexander S. and {Spiridonova}, Olga I.},
        title = "{A massive white-dwarf merger product before final collapse}",
      journal = {\nat},
         year = 2019,
        month = may,
       volume = {569},
       number = {7758},
        pages = {684-687},
          doi = {10.1038/s41586-019-1216-1},
archivePrefix = {arXiv},
       eprint = {1904.00012},
 primaryClass = {astro-ph.SR},
       adsurl = {https://ui.adsabs.harvard.edu/abs/2019Natur.569..684G}
}

@ARTICLE{Tisserand2020,
       author = {{Tisserand}, P. and {Clayton}, G.~C. and {Bessell}, M.~S. and {Welch}, D.~L. and {Kamath}, D. and {Wood}, P.~R. and {Wils}, P. and {Wyrzykowski}, {\L}. and {Mr{\'o}z}, P. and {Udalski}, A.},
        title = "{A plethora of new R Coronae Borealis stars discovered from a dedicated spectroscopic follow-up survey}",
      journal = {\aap},
         year = 2020,
        month = mar,
       volume = {635},
          eid = {A14},
        pages = {A14},
          doi = {10.1051/0004-6361/201834410},
archivePrefix = {arXiv},
       eprint = {1809.01743},
 primaryClass = {astro-ph.SR},
       adsurl = {https://ui.adsabs.harvard.edu/abs/2020A&A...635A..14T}
}

@ARTICLE{Clayton1996,
       author = {{Clayton}, Geoffrey C.},
        title = "{The R Coronae Borealis Stars}",
      journal = {\pasp},
         year = 1996,
        month = mar,
       volume = {108},
        pages = {225},
          doi = {10.1086/133715},
       adsurl = {https://ui.adsabs.harvard.edu/abs/1996PASP..108..225C}
}

@ARTICLE{IbenTutukov1985,
       author = {{Iben}, Jr., I. and {Tutukov}, A.~V.},
        title = "{On the evolution of close binaries with components of initial mass between 3 M and 12 M.}",
      journal = {\apjs},
         year = 1985,
        month = aug,
       volume = {58},
        pages = {661-710},
          doi = {10.1086/191054},
       adsurl = {https://ui.adsabs.harvard.edu/abs/1985ApJS...58..661I}
}

@ARTICLE{Paczynski1971,
       author = {{Paczy{\'n}ski}, B.},
        title = "{Evolutionary Processes in Close Binary Systems}",
      journal = {\araa},
         year = 1971,
        month = jan,
       volume = {9},
        pages = {183},
          doi = {10.1146/annurev.aa.09.090171.001151},
       adsurl = {https://ui.adsabs.harvard.edu/abs/1971ARA&A...9..183P}
}

@ARTICLE{Justham2011,
       author = {{Justham}, Stephen and {Podsiadlowski}, Philipp and {Han}, Zhanwen},
        title = "{On the formation of single and binary helium-rich subdwarf O stars}",
      journal = {\mnras},
         year = 2011,
        month = jan,
       volume = {410},
       number = {2},
        pages = {984-993},
          doi = {10.1111/j.1365-2966.2010.17497.x},
archivePrefix = {arXiv},
       eprint = {1008.1584},
 primaryClass = {astro-ph.SR},
       adsurl = {https://ui.adsabs.harvard.edu/abs/2011MNRAS.410..984J}
}

@ARTICLE{SaioNomoto1998,
       author = {{Saio}, Hideyuki and {Nomoto}, Ken'ichi},
        title = "{Inward Propagation of Nuclear-burning Shells in Merging C-O and He White Dwarfs}",
      journal = {\apj},
         year = 1998,
        month = jun,
       volume = {500},
       number = {1},
        pages = {388-397},
          doi = {10.1086/305696},
archivePrefix = {arXiv},
       eprint = {astro-ph/9801084},
 primaryClass = {astro-ph},
       adsurl = {https://ui.adsabs.harvard.edu/abs/1998ApJ...500..388S}
}

@ARTICLE{Iben1990,
       author = {{Iben}, Jr., Icko},
        title = "{On the Consequences of Low-Mass White Dwarf Mergers}",
      journal = {\apj},
         year = 1990,
        month = apr,
       volume = {353},
        pages = {215},
          doi = {10.1086/168609},
       adsurl = {https://ui.adsabs.harvard.edu/abs/1990ApJ...353..215I}
}

@ARTICLE{vanKerkwijk2010,
       author = {{van Kerkwijk}, Marten H. and {Chang}, Philip and {Justham}, Stephen},
        title = "{Sub-Chandrasekhar White Dwarf Mergers as the Progenitors of Type Ia Supernovae}",
      journal = {\apjl},
         year = 2010,
        month = oct,
       volume = {722},
       number = {2},
        pages = {L157-L161},
          doi = {10.1088/2041-8205/722/2/L157},
archivePrefix = {arXiv},
       eprint = {1006.4391},
 primaryClass = {astro-ph.SR},
       adsurl = {https://ui.adsabs.harvard.edu/abs/2010ApJ...722L.157V}
}

@ARTICLE{Guerrero2004,
       author = {{Guerrero}, J. and {Garc{\'\i}a-Berro}, E. and {Isern}, J.},
        title = "{Smoothed Particle Hydrodynamics simulations  of merging white dwarfs}",
      journal = {\aap},
         year = 2004,
        month = jan,
       volume = {413},
        pages = {257-272},
          doi = {10.1051/0004-6361:20031504},
       adsurl = {https://ui.adsabs.harvard.edu/abs/2004A&A...413..257G}
}

\end{document}